%% file: NekDoubGen.tex
\documentclass[11pt,a4paper]{article}
\usepackage[utf8]{inputenc}
\usepackage[english]{babel}
\usepackage{amsmath}
\usepackage{amsfonts}
\usepackage{amssymb}
\usepackage{graphicx}
\usepackage{color}
\usepackage{transparent}
\usepackage{fullpage}

\date{}
\author{ }
\title{}

\begin{document}

\def\papertitlepage{\baselineskip 3.5ex \thispagestyle{empty}}
\def\preprinumber#1#2{\hfill
\begin{minipage}{1.0in}
#1 \par\noindent #2
\end{minipage}}

\papertitlepage
\preprinumber{YITP-12-30}{} 
\vskip 2ex

\begin{center}

{\LARGE\bf\mathversion{bold}
Double genus expansion for general $\Omega$ background}

\vspace{+30pt}

\end{center}

\baselineskip=3.5ex

\begin{center}
  Andrea Prudenziati\\

{\small
\vskip 3ex
{\it Yukawa Institute for Theoretical Physics, Kyoto University}\\[1ex]
{\it Kyoto 606-8502, Japan}\\
\vskip 1ex
{\tt prude@yukawa.kyoto-u.ac.jp }

}
\end{center}

\baselineskip=3.5ex

\vspace{+10pt}

\begin{abstract}

We will show how the refined holomorphic anomaly equation obeyed by the Nekrasov partition function at generic $\epsilon_1,\epsilon_2$ values becomes compatible, in a certain two parameters expansion, with the assumption that both parameters are associated to genus counting.  The underlying worldsheet theory will be analysed and constrained in various ways, and we will provide both physical interpretation and some alternative evidence for this model. Finally we will use the Gopakumar - Vafa formulation for the refined topological string in order to give a more quantitative description.

\end{abstract}



\vspace{+20pt}

\section{Introduction}

The interpretation of the higher order terms in the $\epsilon_1,\epsilon_2$ expansion of the Nekrasov partition function as gravitational corrections for the gauge theory, has been central in many works since the first conjecture in \cite{Nekrasov:2002qd}. In the limit $\epsilon_{1} = - \epsilon_{2}$ it is now well understood that the underlying microscopic theory is the geometrically engineered topological string; open question is the general case. This hypothetical one parameter refinement of topological strings can be seen either as a deformation of the background in which the theory lives or of the worldsheet 2-d theory itself ( or some combination of the two ). We will concentrate on the latter possibility for which, in the past, various suggestions have already been proposed, \cite{Antoniadis:2010iq} and \cite{Antoniadis:2011hq}, \cite{Huang:2011qx} ( and \cite{Nakayama:2011be} working with heterotic strings ), even if none of them, to my understanding, fully successful. 

The possibility that such a refined theory actually exists and it is some kind of string theory is highly probable essentially because, depending on the A or B model version of the refinement, it shares different and inherently "stringy" features of the topological models, with only a few modifications. A list would include its relationship with the graviphoton supergravity correction \cite{Nekrasov:2002qd}, the connection with the counting of BPS D0-D2 brane bound states \cite{Iqbal:2007ii}, the existence of a refinement of a well established computational method for toric Calabi Yau ( again \cite{Iqbal:2007ii} ) and, central for this paper, the fact that it satisfies a slight deformation of the holomorphic anomaly equation \cite{Huang:2011qx}, \cite{Huang:2010kf},  \cite{Krefl:2010fm}, \cite{Krefl:2010jb} ( and \cite{Krefl:2011aa} for application to ALE space ). This equation has been originally derived from the very worldsheet definition of topological string theory and so it is plausible that its refined version not only suggests the existence of a refined purely worldsheet-type description but it also gives hints for its actual form. 

Following this idea we have found an explicit two parameter expansion for the Nekrasov partition function where the novel deformation parameter that arises away from the $\epsilon_{1} = - \epsilon_{2}$ limit behaves as if being associated to an additional genus expansion, these new handles different from the purely topological ones ( meaning those associated to the parameter that survives the $\epsilon_{1} = - \epsilon_{2}$ limit ). 

In section \ref{sez1} we will give a brief ( and much incomplete ) review for the derivation of the holomorphic anomaly equation emphasizing its relationship with the definition of the topological string amplitudes. 

In section \ref{sez2} it will be (re)explained in what sense the Nekrasov partition function satisfies the holomorphic anomaly equation, what shape this equation assumes depending on the parametrization chosen and what are the consequences.  The main result will be the construction of a general ansatz for the proposed worldsheet definition of the refined string amplitudes satisfying two conditions:

\begin{itemize}
\item It should reduce to the topological string amplitudes in the limit $\epsilon_{1} = - \epsilon_{2}$.
\item It has to be consistent with the refined holomorphic anomaly equation.
\end{itemize}

Section \ref{sez3} tackles the problem of the relationship between the Nekrasov partition function and F-terms in the four dimensional effective field theory, generalizing the well known results of  \cite{Antoniadis:1993ze}. While until this point we have worked on the B model  side, in this section we will switch to the A model perspective\footnote{Throughout the article it will be implicitly assumed that the additional-genus  definition for the refined model is shared by both the A and B model version. Only the B model case is so far computationally justified by the holomorphic anomaly equation, but many issues are more clear and natural if also the A model supports a similar construction. }. The discussion here will be somehow naive, nonetheless the goal is twofold: to provide  physical reason for the two coupling constants expansion of our refined partition function and to explicitly identify inside the Nekrasov partition function what is related to the unconstrained part of our worldsheet ansatz.  

Finally in section  \ref{sez4} we analyse more quantitatively an easy example using the Gopakumar-Vafa interpretation of topological string.

\section{Holomorphic anomaly equation for topological amplitudes}\label{sez1}

Let us start the discussion with the holomorphic anomaly equation \cite{Bershadsky:1993cx} ( \cite{Walcher:2007tp} for the open case with purely closed moduli ). This is a recursion relation for topological strings that equates an antiholomorphic target space moduli derivative of a genus $g$ amplitude to covariant derivatives of lower genus amplitudes. It can be derived starting from the definition of a topological amplitude as an integral over the moduli space of the Riemann surface weighted by a Jacobian given by the contraction of the spin two supercurrents\footnote{I denote $G^{-}$ and $\bar{G}^{-}$ to be the two supercurrents that have been twisted to spin two, while $G^{+}$ and $\bar{G}^{+}$ to spin one. This is the typical B-model notation, for the A-model case the sign of the $U(1)$ current flips in the right moving sector} $G^{-}$ and $\bar{G}^{-}$ with the appropriate Beltrami differentials $\mu_{a}$, $\bar{\mu}_{a}$:

\begin{equation}\label{a}
F^{g} = \int_{{\cal{M}}_{g}}\langle\prod_{a,\bar{a}=1}^{3g-3}\int_{\Sigma_{g}}\hspace{-0,2cm}G^{-}\mu_{a}\int_{\Sigma_{g}}\hspace{-0,2cm}\bar{G}^{-}\bar{\mu}_{a}\rangle_{\Sigma_{g}}
\end{equation}
 
Being $t^{i}$ and $\bar{t}^{\bar{i}}$ the coordinates for the complexified target moduli space of the theory, deriving with respect to $\bar{t}^{\bar{i}}$ corresponds to insert in the amplitude the operator associated to the $\bar{t}^{\bar{i}}$-deformation of the action. This turns out to be exact with respect to the topological charge, $Q = G^+_0 + \bar{G}^+_0$, and so naively trivial. However, before annihilating the vacuum, $G^{+}_0$ and $\bar{G}^{+}_0$ should commute with the $G^{-}_0$ and $\bar{G}^{-}_0$ insertions from (\ref{a}); the result is, because of the supersymmetry algebra of the original worldsheet non linear sigma model, the contraction of the energy momentum tensor with a Beltrami differential which is the same as an external derivative with respect to the Riemann surface moduli:

\[
\bar{\partial}_{\bar{i}} F^{g} = \int_{{\cal{M}}_{g}}\langle\prod_{a,\bar{a}=1}^{3g-3}\int_{\Sigma_{g}}\hspace{-0,2cm}G^{-}\mu_{a}\int_{\Sigma_{g}}\hspace{-0,2cm}\bar{G}^{-}\bar{\mu}_{a}\int_{\Sigma_{g}}\hspace{-0,2cm}d^{2}z\oint_{C_{z}} \hspace{-0,2cm}G^{+}\oint_{C'_{z}}  \hspace{-0,2cm}\bar{G}^{+}\bar{O}_{\bar{i}}(z,\bar{z})\rangle_{\Sigma_{g}} =
\]
\[
=  \int_{{\cal{M}}_{g}}\langle\sum_{b=1}^{3g-3}\partial_{b}\bar{\partial}_{b}\prod_{a,\bar{a}\neq b,\bar{b} }\int_{\Sigma_{g}}\hspace{-0,2cm}G^{-}\mu_{a}\int_{\Sigma_{g}}\hspace{-0,2cm}\bar{G}^{-}\bar{\mu}_{a}\int_{\Sigma_{g}}\hspace{-0,2cm}\bar{O}_{\bar{i}}(z,\bar{z})\rangle_{\Sigma_{g}} 
\]

This gives a nonvanishing contribution only from the ( real codimension two ) boundary of the moduli space of $\Sigma_{g}$ which amounts to a degenerate Riemann surface with either one dividing or non dividing handle shrunk to form a node. With the additional property that the boundary contribution is vanishing unless the $\bar{O}_{\bar{i}}$ sits on the node, and replacing twice the long thin tube on the right and the left of the $\bar{O}_{\bar{i}}$ position, by complete sets of marginal operators plus the appropriate metric, brings the result

\begin{equation}\label{b}
\bar{\partial}_{\bar{i}} F^{g} = \frac{1}{2}\bar{C}_{\bar{i}}^{j k}\left(D_{j}D_{k}F^{g-1} + \sum_{r=1}^{g-1}D_{j}F^{r}D_{k}F^{g-r}  \right)
\end{equation}

The marginal operator insertions are represented by the covariant holomorphic derivatives  whose connection has both a mathematical interpretation on a certain bundle over the moduli space, and a physical one as  regularization for divergent contact terms; the first term on the right of (\ref{b}) corresponds to a non dividing handle degeneration ( two operators on the same $g-1$ Riemann surface ) while the second to a dividing one ( one operator on each genus $g-r$ and $r$ surface ). Finally the sphere three points function  $\bar{C}_{\bar{i}}^{j k}$ is what remains of the portion of the long thin tube on which the $\bar{O}_{\bar{i}}$ and two marginal operators were dwelling.
The story can be generalized for amplitudes with $h$ boundaries ( and crosscaps ) and $m$ operator insertions, $F^{g,h}_{i_1, \dots,i_m}$, still the moral remains the same: the antiholomorphic derivative forces the commutation of the topological currents with part of the measure of the amplitude, thus bringing contributions only from the boundary of the moduli space of Riemann surfaces. In the most general case with purely closed moduli the boundary component can be either of real codimension one or two and, beside the straightforward generalization of (\ref{b}), it contains only the additional terms:

\begin{itemize}
\item $-D_{j}F^{g,h-1}_{j_{1}\dots j_{m}}\Delta_{\bar{i}}^{j}$ corresponding to a boundary shrinking to zero size or, conformally equivalent, moving far away from the rest of the Riemann surface, giving rise to an amplitude with one less  hole and the usual covariant derivative ( $D_{j}F^{g,h-1}_{j_{1}\dots j_{m}}$ ), contracted with a disk containing  $\bar{O}_{\bar{i}}$  and the other marginal operator ( $\Delta_{\bar{i}}^{j}$ ).
\item $-(2g-2+h+m-1)\sum_{s=1}^{m}G_{\bar{i}j_{s}}F^{g,h}_{j_{1}\dots j_{s-1}j_{s+1}\dots j_{m}}$ coming from contact terms arising when $\bar{O}_{\bar{i}}$ approaches one of the operator insertions on a Riemann surface of non vanishing worldsheet curvature.
\end{itemize}

\section{Holomorphic anomaly equation for refined amplitudes}\label{sez2}

In this section we consider the generalization of the holomorphic anomaly equation for the refined topological string amplitudes. First we explain how the antiholomorphic dependence by the target space moduli arises in the Nekrasov partition function. Then we write what is the equation obeyed by these anthiholomorphic moduli using different parameterizations in $\epsilon_{1},\epsilon_{2}$ and selecting one ( parameterization ) for the construction of our worldsheet ansatz. Long time will be spent dealing with the analysis of all the constrains on our ansatz coming from the refined holomorphic anomaly equation. 

\subsection{The antihomorphic dependence of the Nekrasov partition function}

The Nekrasov partition function $Z(\epsilon_{1},\epsilon_{2},a)$ represents a regularized partition function for  supersymmetric $\emph{N}=2$ four dimensional gauge theories, \cite{Nekrasov:2002qd} and \cite{Nekrasov:2003rj}, and depends on the regularization parameters $\epsilon_{1},\epsilon_{2}$, on the vacuum expectation values for the diagonal ( usually $SU(N)$ ) gauge field $a_{1}\dots a_{N-1}$, later on briefly indicated as "$a$", end eventually on various masses ( the instanton counting parameter is kept implicit ). 

We know from the celebrated work of Seiberg and Witten \cite{Seiberg:1994rs} that $Z(\epsilon_{1},\epsilon_{2},a)$ is expected to be holomorphic in the $a$ but for a finite number of singularities. Moreover $a$ is valid only locally as a parameter on the moduli space and objects depending on it will in general transform under the modular group contained in $Sp(2N-2,\mathbb{Z})$, with physical quantities required to be invariant. In order to achieve this latter condition for the coefficients of a generic expansion in the $\epsilon_{1},\epsilon_{2}$ parameters of  $\log Z(\epsilon_{1},\epsilon_{2},a)$, it is in general not possible to maintain the local holomorphic dependence by $a$ but it is necessary to add an antiholomorphic completion, \cite{Huang:2011qx},\cite{Huang:2010kf}, \cite{Krefl:2010fm} and \cite{Krefl:2010jb}. This can be made explicit writing the amplitudes in terms of modular holomorphic functions plus quasi-modular holomorphic generators that become modular after adding an antiholomorphic piece. For example for $SU(2)$ we have the amplitudes depending on the Eisenstein series $E_{2}(\tau)$  \cite{Huang:2011qx},\cite{Huang:2010kf}. Under the redefinition $E_{2}(\tau)\rightarrow \hat{E_{2}}(\tau,\bar{\tau}) = E_{2}(\tau)+6i/\pi (\bar{\tau} - \tau)$, with $\tau=\frac{1}{2\pi i}\frac{\partial^{2}F}{\partial a^{2}}$ and $F$ the prepotential, they become modular invariant objects and all the antiholomorphic dependence is carried by $\hat{E_{2}}(\tau,\bar{\tau})$ . So we should consider the holomorphicity as arising only as a local limit of well defined modular non holomorphic objects. 

This is indeed the same story as for the holomorphic anomaly equation. Chosen a point in the moduli space by fixing $t$ and $\bar{t}$, the topological amplitudes are holomorphic only with respect to local coordinates, usually called $x$, parameterizing marginal operator deformations of the action, \cite{Bershadsky:1993cx}. This $x$ is also the parameter with respect to which the F terms  $F_{g}(x)R_{-}^{2}T_{-}^{2g-2}$, in the four dimensional effective field theory from type II superstrings compactification, are holomorphic. These F terms come from the superspace integration of $F_{g}(X)W^{2g}$, where $F_{g}$ is a function of the chiral superfields $X^{i}$, $W$ is the gravitational superfield and $T_{-}$ and $R_{-}$ were respectively the self dual part of the field strength of the graviton and the graviphoton. They "correspond" to the topological amplitudes in the sense that the coefficient $F_{g}(x)$ is the genus $g$ amplitude for the topological theory with target space the same Calabi Yau used for the superstring compactification \cite{Antoniadis:1993ze}. Being an F term we expect it to be holomorphic in the parameters ( $x$ ) with the holomorphic anomaly equation introducing the correct global $t,\bar{t}$ dependence.

\subsection{The refined holomorphic anomaly equation}

In \cite{Huang:2011qx},\cite{Huang:2010kf}, \cite{Krefl:2010fm} and \cite{Krefl:2010jb}  it has been shown that the coefficients of a two parameters expansion of the Nekrasov partition function satisfy an equation that is but a slight extension of the holomorphic anomaly equation for topological string amplitudes. This can be checked assuming the form of the equation and then solving it order by order to match the Nekrasov results in the holomorphic limit for a variety of examples. 

It was also shown how the holomorphic ambiguity that arises when integrating equations like (\ref{b}), can be completely fixed relying on the behaviour of the Nekrasov partition function in selected points of the moduli space. 

An obvious but important thing to keep in mind is that the explicit form of the equation depends on the chosen expansion. The authors of \cite{Krefl:2010fm} define\footnote{The range of the integer $n$ from $-g$ to $g$ is the one used in all the ansatz inside \cite{Krefl:2010fm} and it is always enough for a full parameterization; in fact it is even overabundant. }.

\begin{equation}\label{d}
\log Z(\epsilon_{1},\epsilon_{2},a)=\sum_{g=0 }^{\infty}\sum_{n=-g }^{g}(-\epsilon_{1}\epsilon_{2})^{g-1}\left(-\frac{\epsilon_{1}}{\epsilon_{2}}\right)^{n}F_{KW}^{g,n}
\end{equation}

Instead \cite{Huang:2010kf} expands as \footnote{the original parameterization has not a minus sign in front of $\epsilon_{1}\epsilon_{2}$ but this brings an "$i$" in the identification with the topological string coupling constant $\lambda$ in the limit $\epsilon_{1}=-\epsilon_{2}=i\lambda$ which is not present in \cite{Krefl:2010fm}. The addition of the minus sign simplifies the comparison and redefines $F_{HK}^{g,n}\rightarrow(-)^{g-1}F_{HK}^{g,n}$. This naively would bring a minus sign on one of the two sides of  (\ref{g}) but, being the three point function $\bar{C}_{\bar{i}\bar{j}\bar{k}}$ derived from the prepotential, it also carries a minus sign and the equation remains untouched. It is easy to check that this agrees completely with (\ref{f})}

\begin{equation}\label{e}
\log Z(\epsilon_{1},\epsilon_{2},a)=\sum_{g,n=0}^{\infty}\left(-\epsilon_{1}\epsilon_{2})^{g-1}(\epsilon_{1}+\epsilon_{2}\right)^{2n}F_{HK}^{g,n}
\end{equation}

In the expansion (\ref{d}) the holomorphic anomaly equation for $F_{KW}^{g,n}$ reads 

\begin{equation}\label{f}
\bar{\partial}_{\bar{i}} F_{KW}^{g}(\beta) = \frac{1}{2}\bar{C}_{\bar{i}}^{j k}(D_{j}D_{k}F_{KW}^{g-1}(\beta) + \sum_{r=1}^{g-1}D_{j}F_{KW}^{r}(\beta)D_{k}F_{KW}^{g-r}(\beta)  )
\end{equation}

where  $F_{KW}^{g}(\beta) = \sum_{n=-g}^{g}(-\epsilon_{1}/\epsilon_{2})^{n}F_{KW}^{g,n}$ and the power of $\beta=-\epsilon_{1}/\epsilon_{2}$ has to be matched on the two sides of the equation. 

Instead using (\ref{e}) we have

\begin{equation}\label{g}
\bar{\partial}_{\bar{i}} F_{HK}^{g,n} = \frac{1}{2}\bar{C}_{\bar{i}}^{j k}(D_{j}D_{k}F_{HK}^{g-1,n} +\hspace{-0,5cm} \sum_{r,s\neq(0,0)\neq(g,n)}D_{j}F_{HK}^{r,s}D_{k}F_{HK}^{g-r,n-s})
\end{equation}

For both (\ref{f}) and (\ref{g}) the first term on the right is zero for $g=0$ and the one loop case $g=1$, that usually has an additional term, is here not different ( due to local Calabi Yau properties ) than the direct $g=1$ reduction of (\ref{f}) and (\ref{g}). Moreover the covariant derivatives appearing are the same as the ones in the usual holomorphic anomaly equation (\ref{b}). Correctly (\ref{f}) and (\ref{g}) are equivalent as can be checked algebraically. 

The partition function can depend on hypermultiplet masses as well. In this case it was found that the holomorphic anomaly equation develops, at least in some examples, an additional term, \cite{Krefl:2010fm}. This has the same form of the degenerating disk that appears in the open holomorphic anomaly equation of \cite{Walcher:2007tp}, and it requires the presence of terms with odd power in $(-\epsilon_{1}\epsilon_{2})^{\frac{1}{2}}$ inside (\ref{d}).  However it was later understood in \cite{Krefl:2010jb} that an appropriate shift for the mass parameters defines a different partition function whose anholomorphic completion obeys the same equation (\ref{g}) of the massless case\footnote{this is similar to the moduli shift of  \cite{Iqbal:2007ii} that brings the refined topological vertex results to have a dependence by $\epsilon_{1}$,$\epsilon_{2}$ compatible to the expansion (\ref{d})}. For this reason from now on we will use (\ref{f}) and (\ref{g}) as the most general cases leaving implicit, when having non zero masses, that the correct shift for them has been performed. 

Let us consider instead 

\begin{equation}\label{h}
\log Z(\epsilon_{1},\epsilon_{2},a)=\sum_{g,n=0}^{\infty}\frac{1}{-\epsilon_{1}\epsilon_{2}}\epsilon_{-}^{2g}\epsilon_{+}^{2n}F^{g,n}  
\end{equation}

with $\epsilon_{\pm}=\frac{\epsilon_{1}\pm\epsilon_{2}}{2}$. In this case  equations (\ref{f}) and (\ref{g}) transform into

\begin{equation}\label{i}
\bar{\partial}_{\bar{i}}F^{g,n} = \frac{1}{2}\bar{C}_{\bar{i}}^{j k}(D_{j}D_{k}F^{g-1,n} -D_{j}D_{k}F^{g,n-1} + \hspace{-0,4cm} \sum_{r,s\neq(0,0)\neq(g,n)}D_{j}F^{r,s}D_{k}F^{g-r,n-s} )
\end{equation}

with the first two terms vanishing if, respectively, $g=0$ and $n=0$.
The main reason behind this work is that in the expansion (\ref{h}) the index $n$ behaves, from the point of view of the above equation, in a way more symmetric with respect to $g$. 

As anticipated it is well known that (\ref{d}), (\ref{e}) and (\ref{h}) all reduce in the limit  $\epsilon_{1} = - \epsilon_{2} = \lambda$ to the expansion 

\[
\log Z \rightarrow \sum_{g=0}^{\infty}\lambda^{2g}F^g  \;\;\;\;\;\; F^g = F^{g,0} = F_{HK}^{g,0} = F_{KW}^{g,0} 
\]

with $F^g$ identified as the topological amplitude of genus $g$ on the local Calabi Yau that geometrically engineers the gauge theory for $Z$. From the left to the right hand side of the arrow it has been performed the holomorphic completion already explained, in this case the dependence by $\bar{t}$ obeys the usual holomorphic anomaly equation (\ref{b}). Needless to say on the left the moduli are gauge theory vevs, on the right geometric objects.

Reversing the derivation that led us to the holomorphic anomaly equation we will start with equation (\ref{i}) as our "experimental data", assume that it is the direct consequence of some yet unknown worldsheet description, and try to derive it.

It is clear that also $F_{KW}^{g,n}$ and $F_{HK}^{g,n}$ can be associated to proper worldsheet amplitudes. In fact the authors of \cite{Huang:2011qx} have already advanced a first proposal for $F_{HK}^{g,n}$ as a genus $g$ topological amplitude with $n$ operator insertions. Given the form of (\ref{g}) and the constraints from the local special geometry they have identified this operator as the first gravitational descendant of the puncture operator, \cite{Witten:1989ig}. However the novelty in the choice of the expansion (\ref{h}) is the peculiar worldsheet interpretation that arises for $F^{g,n}$: as we will see "$n$" no longer counting operators but handles. This new role for the integer $n$ is crucial if we want to completely determine the worldsheet definition for the refined amplitudes, thus fixing also the holomorphic ambiguity left from the constrains derived by (\ref{i}), by comparison with the exact expressions known from direct computation of $\log Z(\epsilon_{1},\epsilon_{2},a)$. We will discuss more later on.

\subsection{Interpretation as a $g+n$ genus amplitude}

Let us for the moment discard the minus sign in front of the second term in the right hand side of (\ref{i}). In this case there is an immediate ansatz for $F^{g,n}$ that explains the form of holomorphic anomaly equation (\ref{i}). $F^{g,n}$ is interpreted as an amplitude of total genus $g+n$ with topological-type measure for the Riemann surface moduli space integral, and "something" marking as different the handles of "type $g$" from the ones of "type $n$" ( which is here expressed by the notation $\langle\dots\rangle_{\Sigma_{g,n}}$ ):

\begin{equation}\label{i11}
F^{g,n} \simeq \int_{{\cal{M}}_{g+n}}\langle\prod_{a,\bar{a}=1}^{3(g+n)-3}\int_{\Sigma_{g+n}}\hspace{-0,5cm}G^{-}\mu_{a}\int_{\Sigma_{g+n}}\hspace{-0,5cm}\bar{G}^{-}\bar{\mu}_{a}\rangle_{\Sigma_{g,n}}
\end{equation}

\begin{figure}[t]
\centering
\vspace{-0pt}
\includegraphics[width=0.8\textwidth]{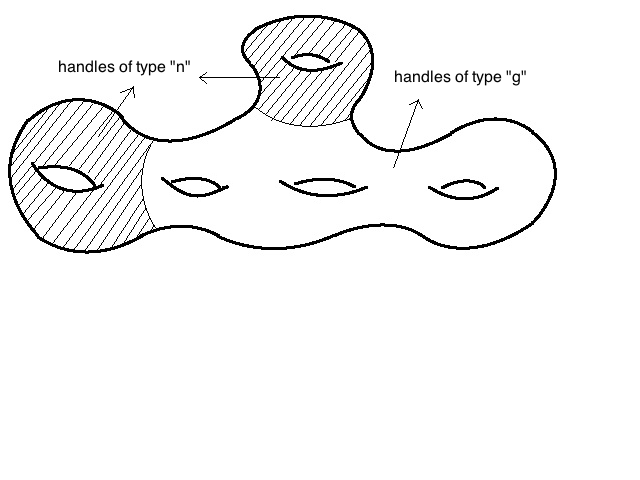}
\vspace{-60pt}
\caption{$F^{g,n}$ is interpreted as a Riemann surface of genus $g+n$, with handles of type $n$ differentiated by the handles of type $g$ in some still undefined way. Pictorially it is shown one of the possible various contributions to $F^{g,n}$ for $g=3$ and $n=2$. The domains containing the handles of type "$n$" are shaded and separated by the domains of type "$g$" by border 1-chains.}
\label{figura1}
\end{figure}

For the moment  we can roughly think to a picture like figure \ref{figura1}, where the Riemann surface is divided into two sets of domains, and the sum of all the genera is $g+n$. In the future we will refer to this situation as "domain decomposition" or "domain splitting". Moreover the amplitude $F^{g,n}$ contains all the possible splittings into domains that satisfy certain geometric conditions we will derive later.  

The computation of $\bar{\partial}_{\bar{i}}F^{g,n}$ proceeds analogously as in section \ref{sez1}, with the topological charge anticommuting with the enlarged moduli space measure of (\ref{i11}), and thus forcing the nonvanishing contributions to come only from the boundary of the moduli space of $\Sigma_{g,n}$. When the degeneration describing these contributions is given by the formation of a node on a non dividing handle belonging to the white "$g$" domain we have a contribution, from that domain decomposition, for the first term in equation (\ref{i}); when it belongs to the shaded "$n$" domain we have a contribution for the second term; and finally when the handle is dividing we obtain contributions for the third term\footnote{a more precise discussion of these issues is in the next section}. 

We want to analyse more in detail the logic that has led us to this first proposal. First of all the integral measure weighting whatever object is associated to the index $n$ still has to contract the Beltrami differentials with the supercurrents $G^{-}$ and $\bar{G}^{-}$ instead that, for example, some ghost field. This is required in order to generate, after the $\bar{\partial}_{\bar{i}}$ derivative, total derivatives with respect to the Riemann surface moduli. The measure for the $g$ handles is further restricted to be exactly the topological string one, as $F^{g,0}$ is required to describe the topological amplitude at genus $g$. However at first sight it would be possible to generalize the "$n$" measure as
 
\[
G^{-}\mu_{a} \rightarrow G^{-}\mu_{a} X \;\;\;with \;\;\; [X_0,G^+_0 + \bar{G}^+_0] = 0 
\]

still generating something proportional to a total derivative for Riemann surface moduli. But, if this was the case, after deforming the topological currents around $G^{-}\mu_{a} X$ and $\bar{G}^{-}\bar{\mu}_{a} \bar{X}$, one $X$ and $\bar{X}$ insertion would remain in the amplitude unpaired to $G^{-}\mu_{a}$ and $\bar{G}^{-}\bar{\mu}_{a}$, and of them we have no trace in the lower $n$-order amplitudes inside equation (\ref{i}).

 The second question is what object should we associate to the integer $n$ in such a way to obtain  the right hand side of (\ref{i})? For a compact Riemann surface there are only four different possibilities: handles, holes, crosscaps and punctures ( with eventually operators ). Crosscaps and holes behave similarly and are in any case connected by the mechanism of tadpole cancellation \cite{Bonelli:2009aw} and \cite{Walcher:2007qp}. As discussed in section \ref{sez1} they degenerate to a single covariant derivative term. Punctures have only the additional term containing the moduli space Kahler metric; a topological amplitude with $n$ appropriate operator insertions is eventually compatible with equation (\ref{g}), not with (\ref{i}). It remains only to consider handles, which fulfill the requirement due to the same degeneration satisfied by $g$ and $n$ ( apart the minus sign ). And this is the reason for the complex dimension of the moduli space to be $3(g+n) -3$.

Four are the important points to develop:

\begin{enumerate}
\item Being now on the same footing $g$ and $n$ it appears that both $\epsilon_{-}$ and $\epsilon_{+}$ inside (\ref{h}) play the role of coupling constants, neither of them being weighted by the complete Euler number but instead sharing it.
\item We need to explain the minus sign in front of the second term on the right hand side of (\ref{i})
\item Some difference between the portion of the Riemann surface counted by $g$ and the one counted by $n$ should appear. First of all in order to distinguish among $g$ and $n$ and among amplitudes with the same $g+n$ value which are not equal in Nekrasov computations.  Second because, if this wasn't the case, the boundary contribution from non dividing handle degeneration to equation (\ref{i}) would come from all the integers $\tilde{n}$ and $\tilde{g}$ such that $\tilde{n} + \tilde{g} = n+g -1$. 
\item We should derive precise geometrical conditions defining the decomposition of the Riemann surface into domains of either $g$ or $n$ phase.
\end{enumerate}

Let us proceed in order and discuss first the apparition of a double coupling constant. This is natural passing to the A-model perspective and using the Gopakumar-Vafa interpretation of topological strings from type II A - M theory, \cite{Gopakumar:1998ii} and \cite{Gopakumar:1998jq}. The coupling constant of the topological partition function is there given by $\lambda=T_{-}g_{s}$ where $T_{-}$ is the vev of the self-dual part of the graviphoton field strength and $g_{s}$ the coupling constant of type II A ( or the radius of M theory ). That case is presently generalized for refined amplitudes allowing also to $T_{+}$ to acquire a vev ( we will discuss this more extensively in section \ref{sez3} ). And in fact we will see that the self-dual and anti self-dual parts of $T$ are respectively $\epsilon_{-}$ and $\epsilon_{+}$. 

For the resolution of the second point we consider two different limits: $\epsilon_{1}=-\epsilon_{2}=\epsilon_{-}$ and $\epsilon_{1}=\epsilon_{2}=\epsilon_{+}$. In the first one the expansion (\ref{h}) reduces to $\sum_{g=0}^{\infty}\epsilon_{-}^{2g-2}F^{g,0}$ and in the second to $\sum_{n=0}^{\infty}-\epsilon_{+}^{2n-2}F^{0,n}$. 
In the first case $F^{g,0}$ is a topological string amplitude of genus $g$ and $\epsilon_{-}$ is the coupling constant\footnote{ a rescaling by the common factor $g_{s}$ is made if seen from the Gopakumar - Vafa point of view }; $F^{0,0}$ plays the role of the prepotential and it corresponds to the sphere. From the previous discussion $-F^{0,n}$ should be thought as some genus $n$ string amplitude with structure similar to (\ref{a}). We have two problems: the first one is the minus sign appearing in the second term in (\ref{i}), when the genus degenerates at the boundary. The second one is the $\epsilon_{+}\rightarrow 0$ limit where  we do not obtain the prepotential, the sphere, but minus it. We then require that it is not $\epsilon_{+}$ to be the coupling constant but instead $\epsilon_{+}/i$. That is we factorize out an "$i$" rewriting the expansion as $\sum_{n=0}^{\infty}(\epsilon_{+}/i)^{2n-2}(i)^{2n}F^{0,n}$ and considering genus $n$ amplitudes the product $(i)^{2n}F^{0,n}$.

This seems a little pedantic but it solves both problems: first  when the new coupling constant goes to zero, $(\epsilon_{+}/i)\rightarrow 0$, we obtain $F^{0,0}$ as it should, and second when we degenerate one non dividing or dividing cycle of "type $n$"  we obtain 

\[
(i)^{2n}F^{0,n}|_{boundary} = (-)^{n}F^{0,n}|_{boundary} \rightarrow
\]
\[
\rightarrow  (i)^{2(n-1)}F^{0,n-1} + \sum_{s=0}^{n}(i)^{2(n-s)}F^{0,n-s}(i)^{2s}F^{0,s} = (-)^{n-1}F^{0,n-1} + \sum_{s=0}^{n}(-)^{n}F^{0,n-s}F^{0,s} 
\]

But equation (\ref{i}) is written in terms of $F^{g,n}$ alone and this brings our minus sign. The same conclusions are valid for non zero $g$ away from any limit on $\epsilon_{1}$ and $\epsilon_{2}$, always identifying Riemann surfaces with $i^{2n}F^{g,n}$. 

\begin{equation}\label{i1}
\sum_{g,n=0}^{\infty}\frac{1}{-\epsilon_{1}\epsilon_{2}}\epsilon_{-}^{2g}\epsilon_{+}^{2n}F^{g,n} = \sum_{g,n=0}^{\infty}\frac{\epsilon_{-}^{2g}(\epsilon_{+}/i)^{2n}}{-\epsilon_{1}\epsilon_{2}}\left[i^{2n}F^{g,n}\right]
\end{equation}

Another consequence of this story is a first $\epsilon_{-}$-$\epsilon_{+}$ or $g$-$n$ asymmetry in the worldsheet definition of our amplitudes as now they are  no longer $F^{g,n}$ but instead $(i)^{2n}F^{g,n}$.  We now move to discuss the third point.

We have seen as the measure for the moduli space integral is fixed to be the one of a purely topological string amplitude of genus $g+n$. So for what matters the coupling to 2-d gravity the handles of type $g$ are equivalent to the handles of type $n$. Similarly it is not possible to distinguish between them by integrating some operators only on the type $n$ domain, as this would not be compatible with equation (\ref{i}) after degeneration of the index $n$. The only possibility left is to make use of the 2-d sigma model action; more specifically we want to glue two different CFTs on the Riemann surface, one with the usual topological Lagrangian $L_{top}$ integrated over the handles of type $g$, the other one with different Lagrangian $L_{top} + \delta L$ integrated over the handles of type $n$. Explicitly:

\begin{equation}\label{v1}
(i)^{2n}F^{g,n} = \int{\emph{D}\phi} \int_{{\cal{M}}_{g+n}} \prod_{a,\bar{a}=1}^{3(g+n)-3}\int_{\Sigma_{g+n}}\hspace{-0,5cm}G^{-}\mu_{a}\int_{\Sigma_{g+n}}\hspace{-0,5cm}\bar{G}^{-}\bar{\mu}_{a} e^{-\int_{\Sigma_{g+n}}\hspace{-0,2cm}L_{top} - \sum_{j}\int_{\Sigma_{n_{j}}}\hspace{-0,2cm} \delta L}
\end{equation}

with $\emph{D}\phi$ a shortcut notation for all the fields in the path integral and $\Sigma_{n_j}$ are all the domains containing each one $n_{j}$ handles of type $n$, $\sum_{j=1} n_j = n$. On the boundary between the two CFTs are assumed in the definition of (\ref{v1}) appropriate boundary conditions for all the fields. 

In order to justify, through a process of gauge fixing of the worldsheet metric, the topological measure also for the handles of type $n$, it is required for $\delta L$ to not spoil the topological behavior of the theory. More in detail a topological charge $Q$ should still be globally conserved on all $\Sigma_{g+n}$ and the supersymmetry algebra has to hold, especially the anticommutators

\[
\{ Q, G^{-}_0\} = T_0  \;\;\; \{ Q, \bar{G}^{-}_0\} = \bar{T}_0
\]

Obviously the explicit realization of $Q$,$G^{-},\bar{G}^{-}$ and $T$ in terms of fields can change passing from a domain to another. Moreover it is necessary that the operators associated to target space antiholomorphic moduli deformations, $\delta\bar{t}^{\bar{i}}\int_{\Sigma_{g}}\hspace{-0,2cm}d^{2}z\oint_{C_{z}} \hspace{-0,2cm}G^{+}\oint_{C'_{z}}  \hspace{-0,2cm}\bar{G}^{+}\bar{O}_{\bar{i}}(z,\bar{z})$, remain the same on the whole Riemann surface, that is for both CFTs: the topological currents are needed for the holomorphic anomaly equation machinery to work and $\bar{O}_{\bar{i}}$ appears on the three point function $\bar{C}_{\bar{i}\bar{j}\bar{k}}$, which is a common factor in the refined holomorphic anomaly equation for both type $g$ and type $n$ degenerations. It is also plausible that operators associated to target space holomorphic moduli deformations $\delta t^i$ should be the same, as the covariant derivatives corresponding to their insertion on the Riemann surface are unchanged from (\ref{b}) to (\ref{i}) and their form has clear physical justification from the form the operator assumes, see \cite{Bershadsky:1993cx}. We will soon give additional evidence for this.  

An important point to note in the definition of our ansatz (\ref{v1}), is that there are not moduli associated to the border along which the two CFTs are glued on the Riemann surface. This means that the 2-d gravity does not distinguish between the two CFTs, the coupling and so the process of gauge fixing being the same. It still can be that the amplitudes $(i)^{2n}F^{g,n}$ modify their value after  deformations of the border keeping unchanged the set of integers $n_{j}$. In the following section it will be shown as this clashes with equation (\ref{i}); thus the picture behind our ansatz is the one of a topological interface between two CFTs.

\subsection{Geometric conditions for the splitting}

We have seen that $(i)^{2n}F^{g,n}$ corresponds to a Riemann surface of genus $g+n$ divided into a certain number of domains either of "type $g$" or "type $n$" defined by the gluing of two different CFTs; one with the usual topological string action, the other a deformation satisfying the above conditions. Moreover the moduli space measure is the one that would appear for a topological string action of genus $g+n$.

We need some rules for the domain decomposition such that the handle degeneration of $(i)^{2n}F^{g,n}$, at the border of the Riemann surface moduli space, is consistent with the results from equation (\ref{i}). Decompositions that do not satisfy these rules should vanish. Presently this is just part of the definition of the ansatz (\ref{v1}), derived from consistency with equation (\ref{i}). To have a microscopical derivation would require the knowledge of $\delta L$ as we will discuss in the conclusions.

\begin{figure}[htb]
\centering
\vspace{+10pt}
\includegraphics[width=0.8\textwidth]{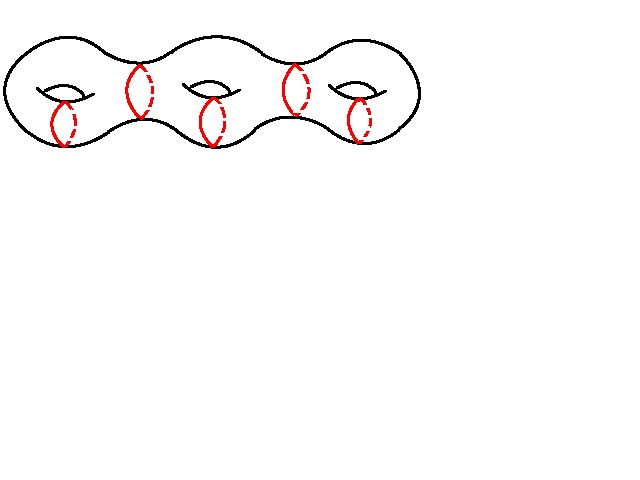}
\vspace{-150pt}
\caption{Chains and cycles that can shrink to form a node on either dividing or non dividing handles.}
\label{figura1a}
\end{figure}

First of all we fix a basis of $g+n$ a-cycles, such that collapsing them corresponds to the shrinking of $g+n$ non dividing handles for the degenerate Riemann surface. There are other non dividing handles that can form a node but for the moment we consider only these. Also we have a set of dividing ( exact ) chains whose collapsing corresponds to a node on a dividing handle. They are represented in figure (\ref{figura1a}). 

\begin{figure}[htb]
\centering
\vspace{+10pt}
\includegraphics[width=0.5\textwidth]{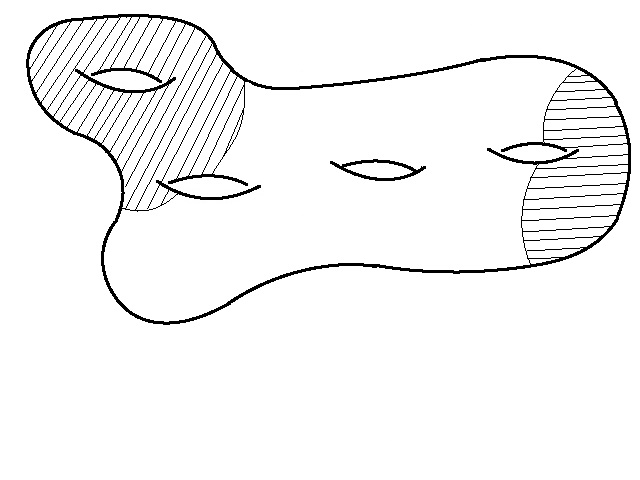}
\vspace{-20pt}
\caption{These are some examples of borders that if you cut along do not give as a result two surfaces whose genus sum is the original total genus. The value of the integers $g$ and $n$ associated is not well defined. The border between any two domains is in this case not connected and they are not allowed in our model}
\label{figura2}
\end{figure}

Let us consider first the case of non dividing handle degeneration, that is we concentrate on the first two terms in equation (\ref{i}).
We want any domain decomposition to belong univoquely to a single $(i)^{2n}F^{g,n}$, in such a way that each moduli space boundary term that arises after the formation of a node ( on a non dividing handle ), will belong either to $(i)^{2n}F^{g-1,n}$ or $(i)^{2(n-1)}F^{g,n-1}$\footnote{ failing this requirement would lead to contributions from amplitudes with the same total number of handles but different values for $g$ and $n$ from the ones appearing on the right hand side of equation (\ref{i}), for example $(i)^{2(n+1)}F^{g-2,n+1}$.}. This is achieved by selecting, for a single $g+n$ Riemann surface, all the domain decompositions such that the sum of all the handles inside all the type $g$ domains is $g$, and the sum of all the handles inside all the type $n$ domains is $n$. Said in another way, when we cut along any border the sum of the genera of the disconnected surfaces so obtained should give the total genus of the original surface. The boundary between any two domains has to be along some connected homologically exact 1-chain. For example figure \ref{figura2} is not allowed. 

\begin{figure}[htb]
\centering
\vspace{+10pt}
\includegraphics[width=0.5\textwidth]{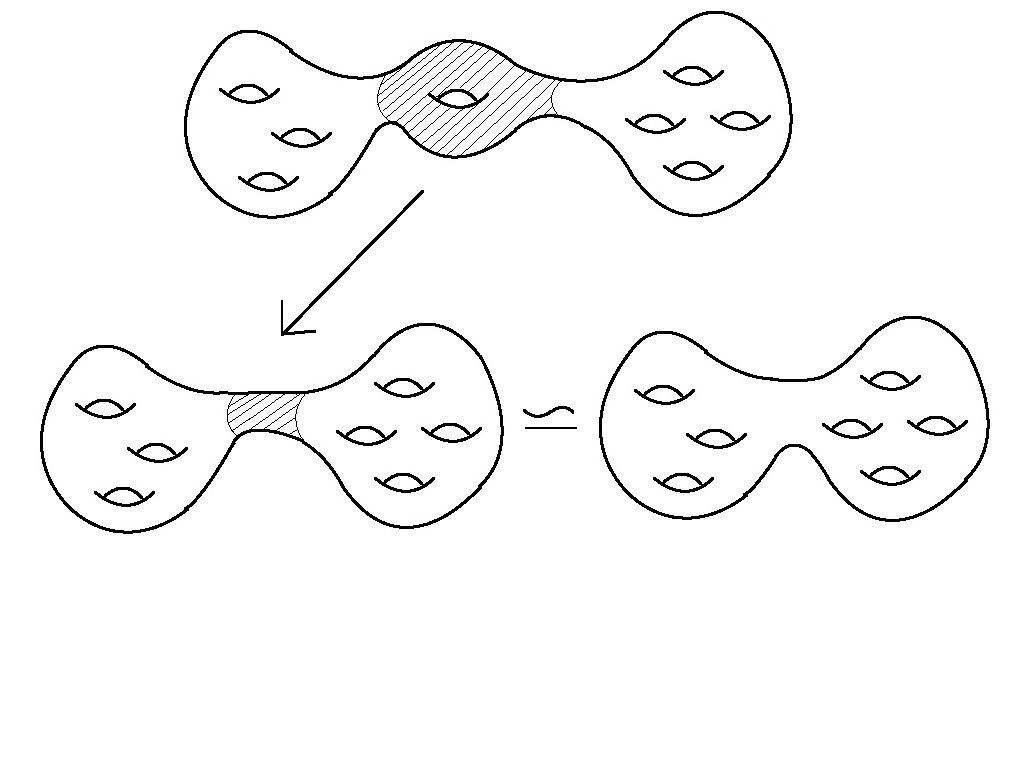}
\vspace{-30pt}
\caption{If a domain is sandwiched between two others ( or more ) it gives rise to and annulus domain ( or generalizations with more borders ) which should be irrelevant as well.}
\label{figura3a}
\end{figure}

\begin{figure}[htb]
\centering
\vspace{+20pt}
\includegraphics[width=0.4\textwidth]{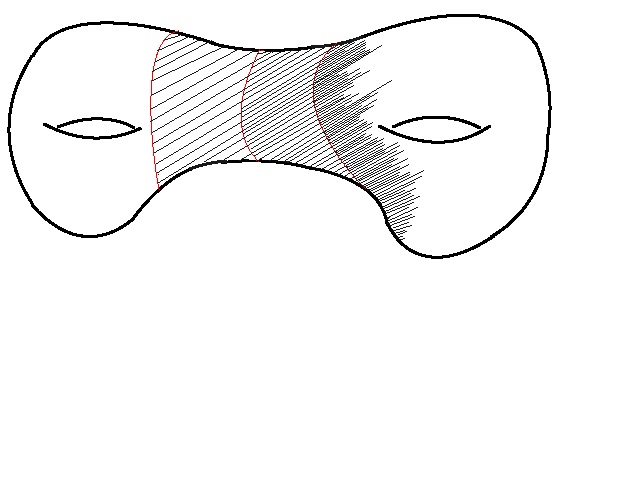}
\vspace{-40pt}
\caption{Three homologically equivalent borders, equivalent also for what matters our model as the region spanned between any two of them has genus zero. The amplitude should remain the same regardless of which one is used}
\label{figura3}
\end{figure}

There is a peculiar case of non dividing handle degeneration that requires additional care,
depicted in figure \ref{figura3a}. If we shrink the non dividing handle of a genus one domain sandwiched among two or more domains in the other phase, we end up with a genus zero domain. We want this to belong to $(i)^{2(n-1)}F^{g,n-1}$ ( or to $(i)^{2n}F^{g-1,n}$ with inverted colors ). This immediately brings us to the requirement depicted in figure \ref{figura3}. 

In figure \ref{figura3}  we are shifting a border between two domains from an exact connected 1-chain to an homologically equivalent one, such that the region spanned by the shift has total genus zero. We ask for this shift to not modify the value of the contribution from that domain decomposition. Also the marginal operator insertions are to be unaffected by the change of domain around their location and so transparent with respect to the above border manipulation\footnote{assuming, as we will justify later, that covariant derivatives in (\ref{i}) correspond to the same usual marginal operator insertions also on domains of type $n$}.
Then the situation in figure (\ref{figura3a}) is simply interpreted as the approaching of the two borders for the central annulus domain, until they coincide and the domain disappears. This is a connected to what we were discussing after equation (\ref{v1}): our ansatz does not have moduli associated to the position and shape of the border between different domains, so from the 2-d gravitational point of view any change as in figure (\ref{figura3}) is irrelevant. Here we are asking that also the purely field theory topological amplitude ( that is before the coupling to 2-d gravity ), does not change under border redefinitions that do not modify the values of the integers $g$ and $n$. 

\begin{figure}[htb]
  \centering
  \vspace{+10pt}
  \def\svgwidth{280pt}
  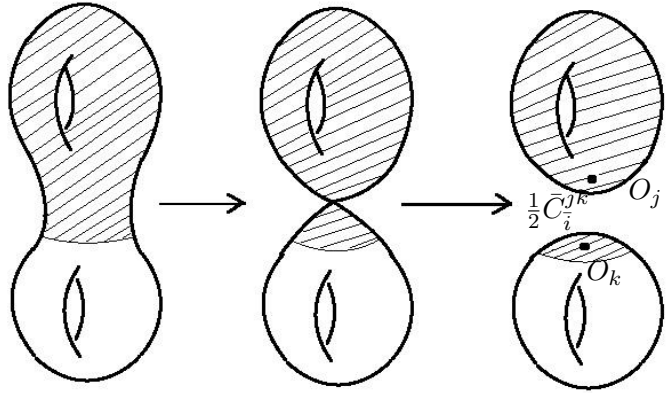
  \vspace{-10pt}  
  \caption{An handle degeneration with the border below the node. After it splits the surface the part below still maintains a disk in the same phase of the part above. From comparison with the holomorphic anomaly equation (\ref{i}) this disk does not alter the value of the amplitude. }
\label{figura4} 
\end{figure}

A similar issue arises when we consider dividing handle degenerations. If the dividing handle we are collapsing belongs to a single domain, then the arising of the corresponding contribution from the last terms in equation (\ref{i}) is immediate. However imagine to look at the situation in figure \ref{figura4}, where along the dividing handle we are degenerating passes also the border between two domains. We can have the node of the degenerating handle above the border, on the border or below. Then, unless the node happened to be exactly on the border 1-chain, we will end up having one of the two pieces with a remaining disk in the other phase with one marginal operator on it. Again this is an application of the condition in figure (\ref{figura3}) with the shaded domain a disk; the border shrinks without affecting the value of the amplitude, until it becomes a point and disappears. This manipulation does not modify the integers $g$ and $n$\footnote{ Note that the fact that domains in the shape of  disks or annuli can appear and disappear leaving the value of the corresponding contribution unchanged is essential also when considering the topological limit $\epsilon_+ \rightarrow 0$ directly applied to the worldsheet ansatz associated to the full partition function. Then this limit can be seen as degenerating all the domains of type $n$ until they reach genus zero, that is they become disks or annuli. The corresponding amplitude is then equivalent to a purely topological one of genus $g$. }.

\begin{figure}[htb]
  \centering
  \vspace{+10pt}
  \def\svgwidth{280pt}
  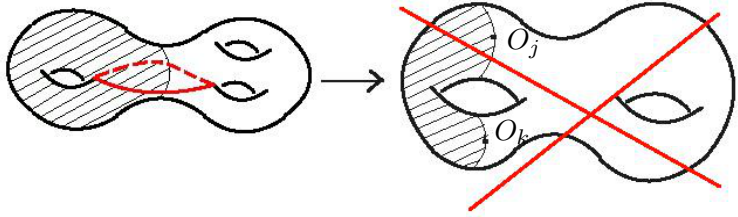
  \vspace{-100pt}  
  \caption{Collapsing the red cycle gives a degenerating non dividing handle with the border passing across longitudinally. The result depicted should vanish. }
\label{figura4a} 
\end{figure}

The last case is the one of a non dividing handle degeneration besides the ones depicted in figure \ref{figura1a}. This is when we have the shrinking of a cycle like the one in figure (\ref{figura4a}). If along this cycle does not pass any border then the boundary contribution belongs to either $(i)^{2n}F^{g-1,n}$ or to $(i)^{2(n-1)}F^{g,n-1}$ in the obvious way, depending on the colors. However if along this handle passes longitudinally the border between two domains, we obtain the result of figure \ref{figura4a}. The domain decomposition on the right hand side of figure \ref{figura4a} is obtained by going at the boundary of the moduli space of the Riemann surface in the domain decomposition represented by the left hand side. The latter does not satisfy our conditions. Thus that handle degeneration and the corresponding boundary contribution is vanishing. This is a first novelty with respect to the usual holomorphic anomaly equation (\ref{b}) for purely topological amplitudes where every handle degeneration contributed. Obviously which non dividing handles contribute or not depends on the splitting we are considering; nonetheless it is clear from figure (\ref{figura1a}) and generalizations that it always exists a basis of $(g+n)$ a-cycles whose collapsing gives non vanishing result.

This concludes the set of requirements which are part of the definition for the domain decomposition appearing in our worldsheet ansatz (\ref{v1}). They have been obtained asking that every possible handle degeneration will give rise only to a contribution appearing inside the amplitudes on the right hand side of equation (\ref{i}).
The next step is to verify the opposite, that is that the most generic object that can appear on the right hand side of equation (\ref{i}) should come only from degeneration of handles from a domain decomposition satisfying the above geometric conditions.

For what matters the last term in equation (\ref{i}) that is quite evident. Every term of the form $D_{j}F^{r,s}D_{k}F^{g-r,n-s}$ is given by two disconnected amplitudes, each one given by the sum of all the allowed domain splitting contributions for integers $r,s$ and $g-r,n-s$, with one marginal operator insertion integrated everywhere. The domain splitting contribution to the bulk amplitude, whose boundary degeneration is of this kind, is simply given by replacing the two operator insertions by an handle. By the stated geometric conditions if the two domains on which the operators are located are of the same type, then the handle belongs to the same domain, otherwise somewhere on it passes the border between them. In both cases the bulk amplitude satisfies by construction our conditions. 

\begin{figure}[htb]
  \centering
  \vspace{+10pt}
  \def\svgwidth{280pt}
  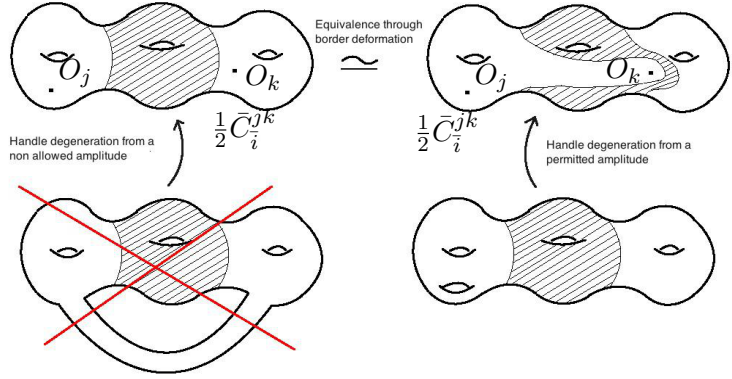
  \vspace{-40pt}  
  \caption{Two equivalent moduli space boundary contributions can be reached by two different moduli space bulk Riemann surfaces. One follows the rules we have discussed, the other does not.}
\label{figura5} 
\end{figure}

It is more tricky the situation coming from the first two terms in equation  (\ref{i}).  Consider for example $D_{j}D_{k}F^{g-1,n}$ ( the other is treated similarly ) that comes from a degenerating non dividing handle of type $g$. The location of the two marginal operators is integrated over all the surface so the degenerating handle they replace could have started and ended from anywhere to anywhere, regardless of the current phases ( $g$ or $n$ ) of the domain on which the marginal operators are. But in the original bulk amplitude both those locations had to belong to the same $g_{i}$ domain, otherwise contradicting the geometric conditions derived. Fortunately, when the two marginal operators sits in points belonging to different domains of the Riemann surface, it exists an equivalent domain decomposition maintaining the same operator positions, but obtained from the non dividing handle degeneration of an allowed amplitude. The border manipulation leading to this has been represented in figure \ref{figura5}. This is a first example of another peculiar fact about our amplitudes. The same component in the boundary of moduli space can be reached from different types of amplitudes in the bulk. In the present case one of the two bulk amplitudes is not allowed and should consequently be zero. But the phenomenon is more general and we will see more later.

\begin{figure}[htb]
  \centering
  \vspace{+10pt}
  \def\svgwidth{680pt}
  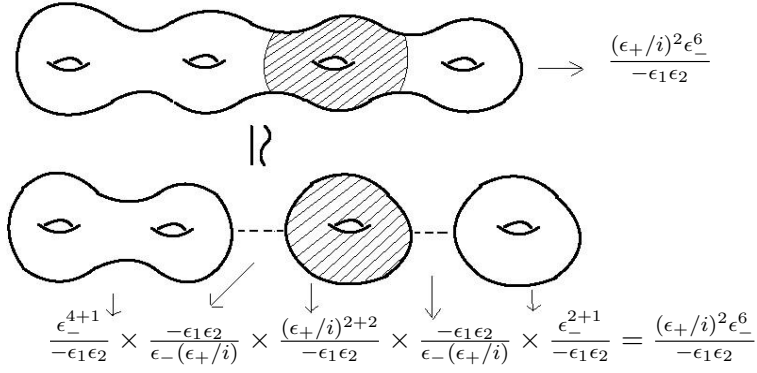
  \vspace{-120pt}  
  \caption{Direct application of the sewing rules with the weights explicitly written}
\label{figura7} 
\end{figure}

It is possible to reconstruct the two coupling constant expansion in (\ref{h}) as a weight coming from a sewing construction from different Riemann surfaces in different phases, correctly weighted. This representation makes clear the string role of $\epsilon_-$ and $\epsilon_+/i$ as coupling constants for the two different CFTs/domains appearing inside our refined topological amplitudes, both arising from the single type IIA dilaton and the two different vevs for the self dual and anti self dual part of the graviphoton. The interpretation is clearly in terms of closed strings either in one phase or the other ( belonging to one type of CFT/domain or the other ): passing across the border is represented as the emission of a string in one phase ( an operator insertion for a generic string state in that phase ) times the propagator given by the inverse of the sphere with two string operator insertions, one in one phase and one in the other, times the absorption by the other domain ( an operator insertion for a generic string state in the other phase ). The rules are easy and represented in figure \ref{figura7} 

\begin{itemize}
\item For each Riemann surface of genus $g_{i}$ ( resp. $n_{j}$) in the corresponding phase that is going to be sewed to $k_{i}$ ( resp. $k_{j}$ ) different surfaces in the $n$ ( resp. $g$ ) phase we multiply by $\epsilon_{-}^{2g_{i}+k_{i}}/(-\epsilon_{1}\epsilon_{2})$ ( resp. $(\epsilon_{+}/i)^{2n_{j}+k_{j}}/(-\epsilon_{1}\epsilon_{2})$ ).
\item For each sewing between two different surfaces we multiply  by the inverse metric $-\epsilon_{1}\epsilon_{2}/((\epsilon_{+}/i)\epsilon_{-})$ ( or the weight of the inverse of a sphere with two operator insertions, one in the $g$ and one in the $n$ phase ).
\item the sewing procedure should create a composite Riemann surface with total genus $g+n$, $g = \sum_{i}g_{i}$ and $n=\sum_{j}n_{j}$. 
\end{itemize}

It can be verified that proceeding in this way the weight does not change adding disks or intermediate annuli or other genus zero domains, thus automatically satisfying our results. Moreover omitting the last rule amounts to reproduce weights that do not fit with the expansion (\ref{h}).  

Last but not least let us look at the enriched moduli space of the new mixed phase Riemann surfaces ( it is exactly the moduli space of genus $g+n$ Riemann surfaces but subdivided into components belonging to different domain decompositions ).  We have seen before as different amplitudes in the bulk of the moduli space ( in that case one of them vanishing ) can degenerate to the same boundary component. The situation is more generic and it involves always non dividing handle degenerations. The prototype case is $\frac{1}{2}\bar{C}_{\bar{i}}^{j k}D_{j}D_{k}(i)^{2n}F^{g,n}$; obviously it can be reached starting from either $(i)^{2n}F^{g+1,n}$ or $(i)^{2n+1}F^{g,n+1}$. We have for this reason that the same boundary component is now shared by two bulk portions of the moduli space or, said in another way, a degenerating non dividing handle does not distinguish between its $g$ or $n$ phase. 

What we have seen in general being true for dividing handles, to be able to change their phase without affecting the value of the amplitude ( the irrelevant appearing or disappearing of an annulus domain from the condition in figure (\ref{figura3}) ), is true for non dividing ones only when degenerating. 

On an infinitely long handle only vacuum states propagates ( as massive states are exponentially suppressed ). Moreover in topological string theory there is a one to one correspondence between the $Q$ cohomology class of operators and differential forms of appropriate degree in the de Rham or Dolbeault cohomology ( in the A or B model respectively ). Then Hodge decomposition applies to corresponding states $|\psi\rangle$

\[
|\psi\rangle = |\psi\rangle_{H} + Q|\chi\rangle
\]

where $|\psi\rangle_{H}$ is annihilated by both $Q$ and $Q^{\dagger}$ and $|\chi\rangle$ is generic. So $|\psi\rangle$ is in the same cohomology class and thus physically equivalent to the vacuum $|\psi\rangle_{H}$, and this defines the correspondence between vacua and physical operators for topological strings. So a long handle can be replaced by a complete set of local topological string operators plus metric, and for conformal invariance to be maintained they also have to be marginal.  But a long degenerating handle spanned by usual topological strings, that is in the $g$ domain, is equivalent to a one in the $n$ domain, as we have seen. Then in both cases it should be replaced by the same set of operator insertions  $\int_{\Sigma_{g}}\hspace{-0,2cm}d^{2}z\oint_{C_{z}} \hspace{-0,2cm}G^{-}\oint_{C'_{z}}  \hspace{-0,2cm}\bar{G}^{-}O_{i}(z,\bar{z})$ thus representing the vacua also in the $n$ domain and so being the operators appearing as substitution for a long handle in the refined holomorphic anomaly equation. Moreover in equation (\ref{i}) the insertion of these operators is represented by a covariant derivative with respect to the target space moduli, so we conclude that the operators associated to a deformation of the holomorphic target space moduli are the same for the action both on the domain of type $g$ and $n$. This means that these operators are in correspondence with either  $H_{d}^{1,1}(X)$ or $H_{\bar{\partial}}^{2,1}(X)$ on all the Riemann surface regardless of the domain to which they belong ( $X$ is the Calabi Yau target space and $d$ and $\bar{\partial}$ refer to the de Rham and Dolbeault cohomology ), and the charge $Q$ maintains the isomorphism with the de Rham or Dolbeault cohomology ( at least ) when restricted to act on marginal physical operators for the usual topological string theory. 

\subsection{Topological antitopological fusion}\label{subsec1}

Before the next section let me clarify a point which was also the original motivation for the present paper. We know that in the limit $\epsilon_{1}=-\epsilon_{2}$ the Nekrasov partition function reduces to the topological one, essentially because both of them can be related to the F terms $R_{-}^{2}T_{-}^{2g-2}$ in the four dimensional  action. Instead if we consider the limit $\epsilon_{1}=\epsilon_{2}$ and we discard for a moment the four dimensional R-symmetry twist which is part of the Nekrasov computation ( see expression (\ref{c}) ), we can repeat the same reasoning only passing from self-dual to anti self-dual objects, $R_{+}^{2}T_{+}^{2n-2}$. But these F terms are now what we would obtain from an antitopological computation, as the bosonized $U(1)$ current in the internal part of the anti self-dual graviphoton vertex operator has an opposite sign with respect to the self-dual case, \cite{Antoniadis:1993ze}. And because this is the piece in the graviton-graviphoton superstring amplitude that resembles the topological twist of the action, we are led to the opposite topological twist, or to an antitopological action. This has now the holomorphic vevs $a$ in the Nekrasov result identified as a local expansion around $\bar{t}$ instead that $t$. So for $(i)^{2n}F^{g,n}$ we would somehow expect a refinement of the topological antitopological fusion construction of \cite{Cecotti:1991me} at higher genus $g+n$, instead of the refinement of our topological genus $g+n$ amplitudes\footnote{the word refinement is here used to refer to the deformation associated to the worldsheet definition of the $n$ domain, be it in topological or antitopological description.}. Apparently this clashes with the holomorphic anomaly equation (\ref{i}) as, passing from a topological part to an antitopological one in the same amplitude, the "anomalous" modulus changes from $\bar{t}$ to $t$, but our equation derives with respect to the lonely $\bar{t}$.  We should note however that in our case the meaning of the holomorphic anomaly equation is: we derive with respect to the moduli that correspond to an anholomorphic completion of the Nekrasov amplitudes, in order to make them modularly invariant. These are $\bar{t}$ for amplitudes holomorphic in $t$, but should be $t$ themselves for amplitudes considered holomorphic in $\bar{t}$. 

The very reason for which antitopological theories are not often considered is that they are completely equivalent to the topological ones, under exchange $t\leftrightarrow\bar{t}$. So we can safely identify the Nekrasov partition function in the limit $\epsilon_{1}=\epsilon_{2}$ ( without the four dimensional R-symmetry twist ) as either the antitopological partition function with $a\rightarrow\bar{t}$ or as the usual topological one with $t$ replacing $\bar{t}$. In both cases the correct answer requires some worldsheet deformation to arise modifying the amplitude, but this is connected to the four dimensional R-symmetry twist we are for the moment omitting. 

\section{Type IIA point of view}\label{sez3}

This section has a double purpose. On one side we want a little more understanding for the role of $\delta L$ inside equation (\ref{v1}). On the other we should discuss why in the refined case we consider a non vanishing vev for both the self dual and anti self dual part of the graviphoton, as required for our double coupling constant interpretation for $\epsilon_-$ and $\epsilon_+/i$ in equation (\ref{i1}).	 To this end we relate here the Nekrasov partition function to certain terms in the four dimensional effective field theory. 

Essentially there are two different possibilities which have been explored in the literature. The first one follows the original idea of Nekrasov to generalize the F terms corresponding to purely genus $g$ topological amplitudes $F_{g}R_{-}^{2}T_{-}^{2g-2} $ to  $F_{g,n}R_{-}^{2}T_{-}^{2g-2}T_{+}^{2n-2}$ inside a not anymore self dual graviphoton background, plus some unknown correction related to a twist in the internal theory, \cite{Nekrasov:2002qd}. The second uses type II - heterotic duality to study a set of refined F terms from compactified heterotic string theory, \cite{Morales:1996bp}. These are $F_{g,n}W^{2g}(\Pi f)$, where $f$ is an arbitrary function of the vector superfield containing the dilaton and $\Pi$ is a chiral projector. It was shown in \cite{Morales:1996bp} that in the large dilaton limit the function $F_{g,n}$ satisfies a refinement of the holomorphic anomaly equation which is exactly our (\ref{i}) without the annoying minus sign on the right side. This looks very promising but it has two problems: the first one is that its worldsheet interpretation constructed in \cite{Antoniadis:2010iq} and \cite{Antoniadis:2011hq} seems not to give the correct results in reproducing the Nekrasov partition function. The second, purely in the spirit of our interpretation, is that it is hard to relate to the philosophy we have followed so far of an $\epsilon_{+}$ - $\epsilon_{-}$ symmetric description plus some deformation. Moreover the dual type II version is somewhat obscure. A priori it would still be possible to associate the value of $\epsilon_{+}$ to some selected vev given by an appropriate combination of the vevs of the vector fields in the dual type II description. However the generalization of the Gopakumar-Vafa idea really relies on an anti self dual vev for the graviphoton, \cite{Gopakumar:1998ii}, \cite{Gopakumar:1998jq} and \cite{Iqbal:2007ii}, so from the present point of view we prefer to follow the original Nekrasov intuition.

In order to do this we write as an index the Nekrasov partition function for rigid $\emph{N}=2$ supersymmetric gauge theories on a five dimensional space $\mathbb{R}^{4}\times\mathbb{S}^{1}$:

\begin{equation}\label{c}
Z(\epsilon_{1},\epsilon_{2},a)=Tr_{\emph{H}_{a}}(-1)^{F}e^{-\beta H}e^{-2\epsilon_{-}J^{3}_{l}}e^{-2\epsilon_{+}\left(J^{3}_{r}+J^{3}_{I}\right)}
\end{equation}

where $\beta$ is the radius of $\mathbb{S}^{1}$, $J_{3}^{l/r/I}$ are the $SU(2)$ Cartan generators respectively for the $SO(4)=SU(2)_{l}\times SU(2)_{r}$ acting on $\mathbb{R}^{4}$ and the internal $SU(2)_{I}$ R-symmetry, and the trace is over the four dimensional Hilbert space after that the gauge field ( generically $SU(N)$ ) has acquired diagonal vacuum expectation values $a_{1}\dots a_{N-1}$.
Already at first sight we can see that the left - right and $\epsilon_{-}$ - $\epsilon_{+}$ symmetry of (\ref{c}) is broken by the single operator $exp(-2\epsilon_{+}J^{3}_{I})$ which in turn is present because of the already mentioned  R-symmetry twist performed during the Nekrasov computation on the four dimensional $\emph{N}=2$ gauge theory, \cite{Nekrasov:2002qd}. The same asymmetry arises in our amplitude interpretation in two fashions: as the $i^{2n}$ appearing in the expansion and as the action deformation $\delta L$ that distinguishes between $n$ and $g$ domains of the full Riemann surface. Moreover as in the limit $\epsilon_{+}=0$ all this disappears leaving us with the correspondence

\begin{equation}\label{l}
\log \left(Tr_{\emph{H}_{a}}(-1)^{F}e^{-\beta H}e^{-2\epsilon_{-}J^{3}_{l}} \right)= \sum_{g=0}^{\infty}\epsilon_{-}^{2g-2}F^{g,0} 
\end{equation}

so in the limit $\epsilon_{-}=0$ we have

\begin{equation}\label{m}
\log \left(Tr_{\emph{H}_{a}}(-1)^{F}e^{-\beta H}e^{-2\epsilon_{+}\left(J^{3}_{r}+J^{3}_{I}\right)}\right)= \sum_{n=0}^{\infty}(\epsilon_{+}/i)^{2n-2}i^{2n}F^{0,n} 
\end{equation}

In this way we isolate both the $i^{2n}$ factorization and the action deformation in the amplitudes $i^{2n}F^{0,n}$ from the purely topological string amplitudes $F^{g,0}$, and we identify them as the appearance on the other side of the equation of the operator $exp(-2\epsilon_{+}J^{3}_{I})$ in the index. In the next section we will give some quantitative measure of this in an easy example. Now we discuss what is the effect of the R-symmetry twist in the four dimensional effective action from the type II A superstring point of view starting from a review of the effect of the $\epsilon_{-}$ $\mathbb{R}^4$ rotation.

If we consider the index in (\ref{c}) in the limit $\epsilon_{+}=0$, we can think the effect of the operator $exp(-2\epsilon_{-}J^{3}_{l} )$ as rotating $\mathbb{R}^4$ while moving around the circle, so obtaining a nontrivial fibration of $\mathbb{R}^4$ over $\mathbb{S}^1$. Then the logarithm of the Nekrasov partition function is equivalent to a vacuum to vacuum amplitude computed on the five dimensional metric called the Omega background:

\begin{equation}\label{n}
ds^2=( dx^{\mu} + \Omega^{\mu}d\theta)^2+d\theta^2
\end{equation}

with $x^{\mu}$ and $\theta$ coordinates along $\mathbb{R}^4$ and $\mathbb{S}^1$ and the value of the (field strength of ) $\Omega^{\mu}$ to be

\begin{equation}\label{o}
T = d\Omega = \epsilon_{1}dx^1\wedge dx^2 + \epsilon_{2}dx^3\wedge dx^4
\end{equation}

Being $\epsilon_{+}=0$, the two form $T$ is self-dual. Reducing the size of $\mathbb{S}^1$ to zero ( $\beta\rightarrow 0$ ) the index localizes around the fixed points of the twist operators and so, for rotations, around the origin. So we can limit ourselves to linear approximation in $x^{\mu}$ in which we can rewrite the metric (\ref{n}) as a Kaluza-Klein ansatz with the Kaluza-Klein gauge field (\ref{o}).

\begin{equation}\label{o1}
ds^2= dx^{2} + ( d\theta + \Omega_{\mu}dx^{\mu})^2
\end{equation}

 Identifying the $\mathbb{S}^1$ of the index with the circle of M theory, geometrically engineered to reproduce the four dimensional gauge theory, the nontrivial Kaluza-Klein background (\ref{o1}) is translated in type II A as a RR two-form self dual field strength background, alternatively referred to as the ( self dual ) graviphoton. If we assume that the coefficients in (\ref{l}) can be identified ( modulo field independent factors ) with higher gravitational corrections in the four dimensional effective action, we should look for F terms that are coupled with the RR two-form self-dual background field. These are the graviphoton F terms $R_{-}^{2}T_{-}^{2g-2}$ already encountered.
 
Actually this reasoning is somehow naive as the identification of (\ref{n}) with (\ref{o1}) is not totally correct even at linear order ( see discussion in \cite{Nakayama:2011be} ). In fact the exact link with topological string amplitudes comes either from direct computations or from type II - heterotic duality and the comparison of the contributions of BPS multiplets on both sides ( again \cite{Nakayama:2011be} and \cite{Gopakumar:1998ii}, \cite{Gopakumar:1998jq} where it is discussed the effect of the additional $R_{-}^{2}$ term to Schwinger computations in purely self dual graviphoton background ). However in order to have some insight about the general case we will simply follow the linear approximation argument also when $\epsilon_{+}\neq 0$. Clearly the results will need future confirmation from more solid arguments.

When $\epsilon_{1}$ and $\epsilon_{2}$ are generic the graviphoton (\ref{o}) is not anymore self dual; further the index in this case contains also an R-symmetry twist, $exp(-2\epsilon_{+}J^{3}_{I})$, that we want to describe in similar way. 

To obtain an R-symmetry rotation on a four dimensional theory ( here only a phase rotation ) we should rotate the supersymmetry generating parameters. Starting from ten dimensional $\emph{N}=2$ supersymmetry we can rotate the chiral spinors for the reduced four dimensional $\emph{N}=2$ by applying an opposite rotation to the covariantly constant spinors of the Calabi Yau compactification. This can be achieved by an isometry on the Calabi Yau ( see for example \cite{Nakayama:2011be} ) that we represent infinitesimally as

\begin{equation}\label{p}
y^{I}\rightarrow y^{I} + v^{I}(y,\epsilon_{+})
\end{equation}

with $y^{I}$ coordinates on the Calabi Yau and the index $I$ running on all the values $i$,$\bar{i}$. Such isometry exists only on non compact Calabi Yau. The $\beta\rightarrow 0$ limit will localize the index around the fixed points of this isometry; if we assume these fixed points isolated we can consider the eleven dimensional background reproducing four dimensional R symmetry plus a ( not anymore self dual ) $\mathbb{R}^4$ rotation, as the fibration of $\mathbb{R}^4$ times the Calabi Yau over $\mathbb{S}^{1}$

\begin{equation}\label{v}
ds^2=( dx^{\mu} + \Omega^{\mu}d\theta)^2 + g_{i\bar{j}}( dy^{i} + v^{i}d\theta)( dy^{\bar{j}} + v^{\bar{j}}d\theta)+d\theta^2
\end{equation}

with $g_{i\bar{j}}$ the internal metric of the Calabi Yau. Now we perform the linear approximation obtaining again a Kaluza-Klein background

\begin{equation}\label{z}
ds^2=dx^{2} + dy_{I}dy^{I}  + (d\theta + \Omega_{\mu}dx^{\mu} + v_{I}dy^{I} )^2
\end{equation}

with the contraction for $y$ performed by $g$. So also this time we reach some RR background in type II A, with part of it in the internal space. 

More precisely: consider type II A geometrically engineered to produce, in the rigid supersymmetry limit, some gauge theory of choice. If we want connection with the Nekrasov partition function with finite generic $\epsilon_1$ and  $\epsilon_2$, this can be seen as a type II A computation looking for higher gravitational corrections involving specific background fields. For  $\epsilon_{+}=0$ this is the self dual graviphoton, so leading to F terms of the kind $R_{-}^{2}T_{-}^{2g-2}$.  For generic values of $\epsilon_1$ and  $\epsilon_2$ we are thinking to amplitudes representing F terms containing both self dual and anti self dual graviphoton fields plus a scalar coming from the Kaluza-Klein four dimensional reduction of the internal RR background  one form $v_{I}$ represented in (\ref{z}). The domains of type $g$ are associated to the self dual graviphoton, the domains of type $n$ to the rest. By comparison the role of $\delta L$ should induce the appearance of the scalar from the K-K reduction of the internal RR background. 

 For a compact Calabi Yau threefold with SU(3) holonomy $h^{0,1}=h^{1,0}=0$, so in this case it is puzzling that, after compactification, a RR one form with internal index does not lead to a four dimensional massless scalar field, but to a massive one. However in geometric engineering we are dealing with a local non compact internal manifold, local limit of a non well specified Calabi Yau $M$. As long as one is interested only in the field theory limit, discarding all the gravitational corrections, the exact form of $M$ is not important; here it is. It would be nice to study in more detail the Kaluza Klein mechanism on these local spaces as perhaps this can explain the puzzling identification of the correct supergravity term corresponding to the omega background, see for example the discussion in \cite{Antoniadis:2010iq}.

\section{Gopakumar-Vafa invariants}\label{sez4}

We conclude our discussion with a recipe for a more quantitative analysis of the problem. More specifically we want to produce explicit numbers to associate to our action deformation; to this end we can look at the Gopakumar-Vafa formulation of the topological A model, \cite{Gopakumar:1998ii} and \cite{Gopakumar:1998jq}.  The GV formulation expresses the topological partition function in term of integers counting the number of D0-D2 branes bound states wrapped on internal Calabi Yau 2-cycles, depending on their four dimensional Lorentz representations. D0 branes naturally couple to RR background one-forms and we have seen how the effect of R-symmetry rotation is represented by this kind of field, so GV invariants should be fully sensitive to our deformation ( even if they are performed in the $\bar{t}\rightarrow\infty$ limit so they will not be sensitive to any $\bar{t}$ effect the $\delta L$ is going to produce ).

\subsection{Refined topological string and GV invariants}

The above mentioned integers are $N^{(j_l,j_r)}_{C}$, corresponding to the number of D0-D2 branes bound states, with the D2 brane wrapping a holomorphic curve in the Calabi Yau $X$, $C\in H_{2}(X,\mathbb{Z})$, and with four dimensional Lorentz quantum numbers $j_{l},j_{r}$ for $SO(4)=SU(2)_{l}\times SU(2)_{r}$. Then a Schwinger one loop integral brings the result

\begin{equation}\label{a1}
\sum_{g}\lambda^{2g-2}F^{g}(t)= \sum_{C\in H_{2}(X,\mathbb{Z})}\sum_{g_{l}}\sum_{k} \frac{\alpha^{g_l}_{C}}{k}(2\sin \frac{k\lambda}{2})^{2g_l -2} e^{-T_{C}k}
\end{equation}

$T_{C}$ is the area of the curve $C$, $\lambda = g_{s}T_{-}$ ( $g_{s}$ is the type II A coupling constant and $T_{-}=\epsilon_{-}$ is the self dual part of the graviphoton vev ). We can rescale on both sides $\lambda\rightarrow \epsilon_{-}$ and work with it. Finally

\begin{equation}\label{b1}
\sum_{g_l}\alpha^{g_l}_{C}I_{l}^{g_l} = \sum_{j_{l},j_r}N^{(j_l,j_r)}_{C}(-1)^{2j_r}(2j_r + 1)[j_l]
\end{equation}

with $[j_l]$ the $SU(2)_{l}$ representation of spin $j_l$ and $I_{l}=(\left[\frac{1}{2}\right] + 2[0])$.
In \cite{Iqbal:2007ii} it has been proposed that for computing refined topological amplitudes it is enough to pass from the coupling to the purely left spin to both right and left, as

\begin{equation}\label{r}
\log Z(t,\epsilon_{1},\epsilon_{2})=\sum_{C\in H_{2}(X,\mathbb{Z})}\sum_{g_{l},g_{r}}\sum_{k} \frac{\alpha^{g_l,g_{r}}_{C}}{k}\frac{(2\sin \frac{k\epsilon_{-}}{2})^{2g_l }(2\sin \frac{k\epsilon_{+}}{2})^{2g_r }}{-4\sin (\frac{k\epsilon_{1}}{2})\sin (\frac{k\epsilon_{2}}{2})}  e^{-T_{C}k}
\end{equation}

with

\begin{equation}\label{s}
\sum_{g_l,g_r}\alpha^{g_l,g_r}_{C}I_{l}^{g_l}\otimes I_{r}^{g_r} = \sum_{j_{l},j_r}N^{(j_l,j_r)}_{C}[j_l]\otimes [j_r]
\end{equation}

What we are going to do is to define certain numbers $\gamma^{g_l,g_r}_{C}$, different from the $\alpha^{g_l,g_r}_{C}$, in term of which we will express our amplitudes stripped by the action deformation, that is reducing to simple $g+n$ topological string amplitudes. Thus a quantitative measure for the effect of $ \delta L$ will be given by the difference between $\alpha^{g_l,g_r}_{C}$ and $\gamma^{g_l,g_r}_{C}$ at each $g_l,g_r$.

We rewrite the partition function for purely $g+n$ topological amplitudes weighted by $\epsilon_{\pm}$ as ( we substitute $F^{g,n}$ with $F^{g+n}$ ) 

\begin{equation}\label{t}
\sum_{g,n}\frac{1}{-\epsilon_{1}\epsilon_{2}}\epsilon_{-}^{2g}\epsilon_{+}^{2n}F^{g+n} = \sum_{C\in H_{2}(X,\mathbb{Z})}\sum_{g_{l},g_{r}}\sum_{k} \frac{\gamma^{g_l,g_{r}}_{C}}{k}\frac{(2\sin \frac{k\epsilon_{-}}{2})^{2g_l }(2\sin \frac{k\epsilon_{+}}{2})^{2g_r }}{-4\sin (\frac{k\epsilon_{1}}{2})\sin (\frac{k\epsilon_{2}}{2})}  e^{-T_{C}k}
\end{equation}

Once again this is our ansatz without any deformation distinguishing between $g$ and $n$ ( apart from the $i^{2n}$ ). Because of (\ref{a1}) we have

\begin{equation}\label{u}
F^{g+n} = i^{-2n}\left[\sum_{C\in H_{2}(X,\mathbb{Z})}\sum_{g_{l}}\sum_{k} \frac{\alpha^{g_l}_{C}}{k}(2\sin \frac{k\epsilon_{-}}{2})^{2g_l -2} e^{-T_{C}k}\right]_{coefficient \;of \;\epsilon_{-}^{2g+2n-2}}
\end{equation}

so we can express the $\gamma^{g_l,g_r}_{C}$ in terms of the $\alpha^{g_l}_{C}$ and then the $N^{(j_l,j_r)}_{C}$ and finally the $\alpha^{g_l,g_r}_{C}$. 

Before giving some numbers in an example for the simple $\mathbb{P}^{1}\times\mathbb{P}^{1}$ case, let me point out the situation. Working purely in terms of type II A - M theory we can compute the refined amplitudes by the knowledge of the numbers $N^{(j_l,j_r)}_{C}$. The only difference between the topological partition function and the refined one is the coupling or not to the right $SU(2)_{r}$ spin, through the appearance of a non vanishing vev for the anti self-dual part of the graviphoton, $\epsilon_{+}$. When it is zero we immediately reduce to the usual topological case. So to speak all the information about the Nekrasov partition function is already contained in the Gopakumar-Vafa invariants $N^{(j_l,j_r)}_{C}$. The choice between computing topological amplitudes or refined ones translates in not coupling or coupling to the right spin.

 From the present point of view instead we start directly from topological amplitudes; in this case switching on the vev for the anti self dual graviphoton implies the addition of a new expansion parameter in the refined topological partition function, counting additional handles. But unlike the type II A case this is not enough. We need to consider some additional deformation for the action integrated on the domain of type $n$, as we have seen. 

\subsection{$\mathbb{P}^{1}\times\mathbb{P}^{1}$}

As an example we consider the local $\mathbb{P}^{1}\times\mathbb{P}^{1}$. Once the $N^{(j_l,j_r)}_{C}$ are given for some $C$, then it is just an algebraic exercise to construct the $\alpha^{g_l,g_r}_{C}$ and derive the $\gamma^{g_l,g_r}_{C}$ regardless of the original geometric origin.
A basis for $C\in H_{2}(\mathbb{P}^{1}\times\mathbb{P}^{1},\mathbb{Z})$ is given by $B$ and $F$ ( base and fiber ) with intersection numbers $B\cdot B = F\cdot F =0$ and $B\cdot F =1$. In the following examples we list the values of  $\alpha^{g_l,g_r}_{C}$ and $\gamma^{g_l,g_r}_{C}$ for various curves ( computation has been performed also for the curves $3B + 3F$ and $3B + 4F$ but omitted for reasons of space ). Inside the tables $g_l,g_r=0,1,2,\dots$ span respectively rows ( the lines ) and columns ( the position on the line ).

\vspace{+20pt}
\begin{tabular}{||l||l||}

\hline
$\gamma^{g_l,g_r}_{B}$ & $\alpha^{g_l,g_r}_{B}$ \\
\hline
\hline
-2, 1/3, 0, 1/1512, 1/9072, 1/49896, 193/50295168, \dots &  -2,1, 0, 0, \dots \\
  0, 1/360, 0, 1/362880, 1/2177280, 21647/261534873600, \dots &  0, 0, \dots \\
0, 1/1512, 0, 1/1330560, 1/7983360, \dots & \dots  \\
 0, 41/362880, 0, 16693/130767436800, \dots & \\
 0, 491/23950080, 0,\dots & \\
0, 341749/87178291200, \dots & \\
 0, \dots & \\
\hline

\end{tabular}
\vspace{-0pt}

\vspace{+10pt}
\begin{tabular}{||l||l||}

\hline
$\gamma^{g_l,g_r}_{B+F}$ & $\alpha^{g_l,g_r}_{B+F}$ \\
\hline\hline
-4, 2/3, 0, 1/756, 1/4536, 1/24948, 193/25147584, \dots &  -4, 10, -6, 1, 0, 0, \dots \\
 0, 1/180, 0, 1/181440, 1/1088640, 21647/130767436800, \dots &  0, 0, \dots \\
 0, 1/756, 0, 1/665280, 1/3991680, \dots & \dots  \\
  0, 41/181440, 0, 16693/65383718400, \dots & \\
 0, 491/11975040, 0, \dots & \\
 0, 341749/43589145600, \dots & \\
 0, \dots & \\
\hline
\end{tabular}
\vspace{-0pt}

\vspace{+10pt}
 \begin{tabular}{||l||l||}
\hline
$\gamma^{g_l,g_r}_{B+2F}$ & $\alpha^{g_l,g_r}_{B+2F}$ \\
\hline\hline
 -6, 1, 0, 1/504, 1/3024, 1/16632, 193/16765056\dots & -6, 35, -56, 36, -10, 1, 0, 0,  \dots \\
0, 1/120, 0, 1/120960, 1/725760, 21647/87178291200, \dots & 0, 0, \dots \\
0, 1/504, 0, 1/443520, 1/2661120, \dots & \dots  \\
 0, 41/120960, 0, 16693/43589145600, \dots & \\
 0, 491/7983360, 0, \dots & \\
0, 341749/29059430400,\dots & \\
0, \dots & \\
\hline
\end{tabular}
\vspace{-0pt}

 \vspace{+10pt}
  \begin{tabular}{||l||l||}
\hline
$\gamma^{g_l,g_r}_{B+3F}$ & $\alpha^{g_l,g_r}_{B+3F}$ \\
\hline\hline
-8, 4/3, 0, 1/378, 1/2268, 1/12474, 193/12573792, \dots &  -8, 84, -252, 330, -220, 78, -14, 1,0, \dots \\
 0, 1/90, 0, 1/90720, 1/544320, 21647/65383718400, \dots & 0, 0, \dots \\
0, 1/378, 0, 1/332640, 1/1995840, \dots & \dots  \\
0, 41/90720, 0, 16693/32691859200, \dots & \\
 0, 491/5987520, 0, \dots & \\
 0, 341749/21794572800,\dots & \\
 0,\dots & \\
\hline
\end{tabular}
\vspace{-0pt}

 \vspace{+10pt}
   \begin{tabular}{||l||}
\hline
$\gamma^{g_l,g_r}_{2B+2F}$ \\
\hline\hline
-32, -11/3, 0, 2/189, 1/567, 2/6237, 193/3143448,\dots \\
9, 2/45, 0, 1/22680, 1/136080, 21647/16345929600, \dots \\
0, 2/189, 0, 1/83160, 1/498960, \dots \\
 0, 41/22680, 0, 16693/8172964800,  \dots \\
  0, 491/1496880, 0, \dots \\
 0, 341749/5448643200,\dots  \\
0 \dots  \\
\hline\hline
 $\alpha^{g_l,g_r}_{2B+2F}$ \\
 \hline\hline
  -32, 359, -1232, 1950, -1660, 807, -224, 33, -2, 0, 0,  \dots \\
  9, -120, 462, -792, 715, -364, 105, -16, 1, 0, \dots \\
 0,0, \dots  \\
\hline
\end{tabular}
\vspace{-0pt}
\vspace{+20pt}

To compare the $\gamma^{g_l,g_r}_{C}$ with the $\alpha^{g_l}_{C}$, these matrices can be roughly divided into three parts.

 The first column, which is equal on both sides. This is not a surprise as the first column refers to the case $n=0$ where we limit to topological amplitudes both for (\ref{r}) and for (\ref{t}),(\ref{u}). 

All the columns from a certain number ( depending on the curve ) to infinity and the lower part, from a certain line to infinity, of the remaining columns ( apart from the very first one ). There the $\alpha^{g_l}_{C}$ are zero and this is an issue of the basis (\ref{s}). For what matters the $\gamma^{g_l,g_r}_{C}$ they are not zero, but they reduce to very small numbers. So the action deformation $\delta L$ is somehow irrelevant in this zone.

The rest, the upper left part of the matrix, apart from the first column. There the difference between $\alpha^{g_l}_{C}$  and the $\gamma^{g_l,g_r}_{C}$ is the biggest.

Finally note that in general $\gamma^{g_l,g_r}_{C}$ is not an integer. So there cannot be interpretation as counting any object.

\section{Conclusions}

In this paper we have constructed a general ansatz for a worldsheet definition of the coefficients of the Nekrasov partition function expanded with respect to the parameters $\epsilon_-$ and $\epsilon_+/i$. 

This ansatz has the property to reduce to the usual genus $g$ topological amplitudes in the limit $\epsilon_+ \rightarrow 0$  and to satisfy the refined holomorphic anomaly equation that is known to held from direct computations. Moreover we have given some physical understanding for the interpretation of  both $\epsilon_-$ and $\epsilon_+/i$ as coupling constant for the refined topological theory.

Finally we have analysed the quantitative discrepancy associated to the effect of $\delta L$ inside equation (\ref{v1}) by comparison with the correct answer obtained from the refined topological vertex.

What is left for transforming this ansatz into an actual uniquely defined worldsheet definition is the direct knowledge of $\delta L$, that is of the CFT associated to the type $n$ domain. It is clear that $\delta L$ is nothing but the worldsheet counterpart of the holomorphic ambiguity from the integration of the refined holomorphic anomaly equation. Thus its knowledge should be obtained by other means. Indeed a possibility for studying this new conformal theory is to consider amplitudes in a unique type $n$ domain, for example $-F^{0,1}$. This from our point of view is nothing but a genus 1 amplitude for the unknown CFT with Lagrangian $L_{top} + \delta L$ for which, in explicit examples, are given exact expressions ( see for example \cite{Nakajima:2003uh} ). Then the goal would be to derive what CFT Lagrangian would give rise to such torus amplitude on the local Calabi Yau considered. Much probably this will need the introduction of new degrees of freedom integrated only on the domains of type $n$ and with correct boundary conditions \footnote{ which unfortunately can be derived rigorously only from, at least, results from genus 2 amplitudes}

An interesting check for our proposal would be to study the theory around the conifold point in moduli space. It is known that there the topological string reduces to the $c=1$ string theory at the self dual radius, \cite{Ghoshal:1995wm}, and the refined case generalizes  multiplying the radius by the ratio between $\epsilon_1$ and $\epsilon_2$. Thus in our case $\delta L$ would have to correspond to the operator deforming the compactification radius of the $c=1$ theory, if restricted to the conifold point in moduli space.

A direct relationship between our proposal and the one given by Huang and Klemm in  \cite{Huang:2011qx} is also interesting. Roughly speaking it is possible that each dilaton operator of \cite{Huang:2011qx} somehow pops up, in our model, to a genus one domain of type $n$, and higher domains correspond to the case of colliding operators. But being the definitions associated to different objects ( even if linearly related one to the other ), the dictionary between the two proposal can be more complicated.

Finally it is auspicable a direct microscopical derivation of our results. Doing this would probably require the knowledge of some 2-d worldsheet theory reducing in some limit to two different phases described by the two CFTs on type $g$ and $n$ domains, together with the geometrical conditions for the domain decomposition. To my knowledge something similar has been done only in \cite{Ooguri:2002gx}.

\subsection*{Acknowledgments}

It is a pleasure to thank Giulio Bonelli, Tohru Eguchi, Kazuhiro Sakai, Alessandro Tanzini and Johannes Walcher for various discussions and useful comments, and Yu Nakayama for email correspondence. Also I would like to thank the CERN Theory Group for hospitality during which I started to tackle the problem, long time ago. This work was supported by the Japanese Society for the Promotion of Science (JSPS).

\end{document}

%% file: sup4n.pdf_tex
\begingroup%
  \makeatletter%
  \providecommand\color[2][]{%
    \errmessage{(Inkscape) Color is used for the text in Inkscape, but the package 'color.sty' is not loaded}%
    \renewcommand\color[2][]{}%
  }%
  \providecommand\transparent[1]{%
    \errmessage{(Inkscape) Transparency is used (non-zero) for the text in Inkscape, but the package 'transparent.sty' is not loaded}%
    \renewcommand\transparent[1]{}%
  }%
  \providecommand\rotatebox[2]{#2}%
  \ifx\svgwidth\undefined%
    \setlength{\unitlength}{552.48326111bp}%
    \ifx\svgscale\undefined%
      \relax%
    \else%
      \setlength{\unitlength}{\unitlength * \real{\svgscale}}%
    \fi%
  \else%
    \setlength{\unitlength}{\svgwidth}%
  \fi%
  \global\let\svgwidth\undefined%
  \global\let\svgscale\undefined%
  \makeatother%
  \begin{picture}(1,0.69504368)%
    \put(0,0){\includegraphics[width=\unitlength]{sup4n.pdf}}%
    \put(0.798769,0.27328382){\color[rgb]{0,0,0}\makebox(0,0)[lb]{\smash{$O_{k}$}}}%
    \put(0.85467408,0.38276456){\color[rgb]{0,0,0}\makebox(0,0)[lb]{\smash{$O_{j}$}}}%
    \put(0.70326448,0.35714139){\color[rgb]{0,0,0}\makebox(0,0)[lb]{\smash{   $\frac{1}{2}\bar{C}_{\bar{i}}^{jk}$}}}%
  \end{picture}%
\endgroup%

%% file: figura4a.pdf_tex
\begingroup%
  \makeatletter%
  \providecommand\color[2][]{%
    \errmessage{(Inkscape) Color is used for the text in Inkscape, but the package 'color.sty' is not loaded}%
    \renewcommand\color[2][]{}%
  }%
  \providecommand\transparent[1]{%
    \errmessage{(Inkscape) Transparency is used (non-zero) for the text in Inkscape, but the package 'transparent.sty' is not loaded}%
    \renewcommand\transparent[1]{}%
  }%
  \providecommand\rotatebox[2]{#2}%
  \ifx\svgwidth\undefined%
    \setlength{\unitlength}{512bp}%
    \ifx\svgscale\undefined%
      \relax%
    \else%
      \setlength{\unitlength}{\unitlength * \real{\svgscale}}%
    \fi%
  \else%
    \setlength{\unitlength}{\svgwidth}%
  \fi%
  \global\let\svgwidth\undefined%
  \global\let\svgscale\undefined%
  \makeatother%
  \begin{picture}(1,0.75)%
    \put(0,0){\includegraphics[width=\unitlength]{figura4a.pdf}}%
    \put(0.68030972,0.65542036){\color[rgb]{0,0,0}\makebox(0,0)[lb]{\smash{$O_j$}}}%
    \put(0.66371679,0.53373895){\color[rgb]{0,0,0}\makebox(0,0)[lb]{\smash{$O_k$}}}%
  \end{picture}%
\endgroup%

%% file: sup5n.pdf_tex
\begingroup%
  \makeatletter%
  \providecommand\color[2][]{%
    \errmessage{(Inkscape) Color is used for the text in Inkscape, but the package 'color.sty' is not loaded}%
    \renewcommand\color[2][]{}%
  }%
  \providecommand\transparent[1]{%
    \errmessage{(Inkscape) Transparency is used (non-zero) for the text in Inkscape, but the package 'transparent.sty' is not loaded}%
    \renewcommand\transparent[1]{}%
  }%
  \providecommand\rotatebox[2]{#2}%
  \ifx\svgwidth\undefined%
    \setlength{\unitlength}{819.2bp}%
    \ifx\svgscale\undefined%
      \relax%
    \else%
      \setlength{\unitlength}{\unitlength * \real{\svgscale}}%
    \fi%
  \else%
    \setlength{\unitlength}{\svgwidth}%
  \fi%
  \global\let\svgwidth\undefined%
  \global\let\svgscale\undefined%
  \makeatother%
  \begin{picture}(1,0.75)%
    \put(0,0){\includegraphics[width=\unitlength]{sup5n.pdf}}%
    \put(0.08191999,0.61546704){\color[rgb]{0,0,0}\makebox(0,0)[lb]{\smash{$O_{j}$}}}%
    \put(0.33167955,0.60758683){\color[rgb]{0,0,0}\makebox(0,0)[lb]{\smash{$O_{k}$}}}%
    \put(0.63880502,0.6091578){\color[rgb]{0,0,0}\makebox(0,0)[lb]{\smash{$O_{j}$}}}%
    \put(0.81396861,0.61701268){\color[rgb]{0,0,0}\makebox(0,0)[lb]{\smash{$O_{k}$}}}%
    \put(0.28690678,0.53689298){\color[rgb]{0,0,0}\makebox(0,0)[lb]{\smash{$\frac{1}{2}\bar{C}_{\bar{i}}^{jk}$}}}%
    \put(0.55789982,0.53218007){\color[rgb]{0,0,0}\makebox(0,0)[lb]{\smash{$\frac{1}{2}\bar{C}_{\bar{i}}^{jk}$}}}%
  \end{picture}%
\endgroup%

%% file: sup7n.pdf_tex
\begingroup%
  \makeatletter%
  \providecommand\color[2][]{%
    \errmessage{(Inkscape) Color is used for the text in Inkscape, but the package 'color.sty' is not loaded}%
    \renewcommand\color[2][]{}%
  }%
  \providecommand\transparent[1]{%
    \errmessage{(Inkscape) Transparency is used (non-zero) for the text in Inkscape, but the package 'transparent.sty' is not loaded}%
    \renewcommand\transparent[1]{}%
  }%
  \providecommand\rotatebox[2]{#2}%
  \ifx\svgwidth\undefined%
    \setlength{\unitlength}{1541.78973999bp}%
    \ifx\svgscale\undefined%
      \relax%
    \else%
      \setlength{\unitlength}{\unitlength * \real{\svgscale}}%
    \fi%
  \else%
    \setlength{\unitlength}{\svgwidth}%
  \fi%
  \global\let\svgwidth\undefined%
  \global\let\svgscale\undefined%
  \makeatother%
  \begin{picture}(1,0.39849792)%
    \put(0,0){\includegraphics[width=\unitlength]{sup7n.pdf}}%
    \put(0.33926838,0.3564958){\color[rgb]{0,0,0}\makebox(0,0)[lb]{\smash{$\frac{(\epsilon_{+}/i)^2\epsilon_{-}^6}{-\epsilon_1\epsilon_2}$}}}%
    \put(0.03069985,0.20134609){\color[rgb]{0,0,0}\makebox(0,0)[lb]{\smash{$\frac{\epsilon_{-}^{4+1}}{-\epsilon_1\epsilon_2} \times \frac{-\epsilon_1\epsilon_2}{\epsilon_{-}(\epsilon_{+}/i)} \times \frac{(\epsilon_{+}/i)^{2+2}}{-\epsilon_1\epsilon_2} \times \frac{-\epsilon_1\epsilon_2}{\epsilon_{-}(\epsilon_{+}/i)} \times \frac{\epsilon_{-}^{2+1}}{-\epsilon_1\epsilon_2} = \frac{(\epsilon_{+}/i)^2\epsilon_{-}^6}{-\epsilon_1\epsilon_2}$}}}%
  \end{picture}%
\endgroup%

%% file: NekDoubGen.bbl
\begin{thebibliography}{20}
 
\bibitem{Antoniadis:1993ze}
  I.~Antoniadis, E.~Gava, K.~S.~Narain and T.~R.~Taylor,
  ``Topological amplitudes in string theory,''
  Nucl.\ Phys.\  B {\bf 413} (1994) 162
  [arXiv:hep-th/9307158].
  
\bibitem{Antoniadis:2010iq}
  I.~Antoniadis, S.~Hohenegger, K.~S.~Narain and T.~R.~Taylor,
  ``Deformed Topological Partition Function and Nekrasov Backgrounds,''
  Nucl.\ Phys.\  B {\bf 838} (2010) 253
  [arXiv:1003.2832 [hep-th]].
  
\bibitem{Antoniadis:2011hq}
  I.~Antoniadis, S.~Hohenegger, K.~S.~Narain and E.~Sokatchev,
  ``Generalized N=2 Topological Amplitudes and Holomorphic Anomaly Equation,''
  Nucl.\ Phys.\  B {\bf 856} (2012) 360
  [arXiv:1107.0303 [hep-th]].

\bibitem{Bershadsky:1993cx}
  M.~Bershadsky, S.~Cecotti, H.~Ooguri and C.~Vafa,
  ``Kodaira-Spencer theory of gravity and exact results for quantum string
  amplitudes,''
  Commun.\ Math.\ Phys.\  {\bf 165} (1994) 311
  [arXiv:hep-th/9309140].

\bibitem{Bonelli:2009aw}
  G.~Bonelli, A.~Prudenziati, A.~Tanzini and J.~Yang,
  JHEP {\bf 0906} (2009) 046
  [arXiv:0905.1286 [hep-th]].
  
\bibitem{Cecotti:1991me}
  S.~Cecotti and C.~Vafa,
  ``Topological antitopological fusion,''
  Nucl.\ Phys.\  B {\bf 367} (1991) 359.
  
\bibitem{Ghoshal:1995wm}
  D.~Ghoshal and C.~Vafa,
  ``c = 1 string as the topological theory of the conifold,''
  Nucl.\ Phys.\ B {\bf 453} (1995) 121
  [hep-th/9506122].

\bibitem{Gopakumar:1998ii}
  R.~Gopakumar and C.~Vafa,
  ``M-theory and topological strings. I,''
  arXiv:hep-th/9809187. 
  
\bibitem{Gopakumar:1998jq}
  R.~Gopakumar and C.~Vafa,
  ``M-theory and topological strings. II,''
  arXiv:hep-th/9812127.

\bibitem{Huang:2011qx}
  M.~x.~Huang, A.~K.~Kashani-Poor and A.~Klemm,
 ``The Omega deformed B-model for rigid N=2 theories,''
  arXiv:1109.5728 [hep-th].

\bibitem{Huang:2010kf}
  M.~x.~Huang and A.~Klemm,
  ``Direct integration for general Omega backgrounds,''
  arXiv:1009.1126 [hep-th].

\bibitem{Iqbal:2007ii}
  A.~Iqbal, C.~Kozcaz and C.~Vafa,
  ``The refined topological vertex,''
  JHEP {\bf 0910} (2009) 069
  [arXiv:hep-th/0701156].

\bibitem{Krefl:2011aa}
  D.~Krefl and S.~-Y.~D.~Shih,
  arXiv:1112.2718 [hep-th].
  
  \bibitem{Krefl:2010fm}
  D.~Krefl and J.~Walcher,
  ``Extended Holomorphic Anomaly in Gauge Theory,''
  Lett.\ Math.\ Phys.\  {\bf 95} (2011) 67
  [arXiv:1007.0263 [hep-th]].
  
\bibitem{Krefl:2010jb}
  D.~Krefl and J.~Walcher,
  ``Shift versus Extension in Refined Partition Functions,''
  arXiv:1010.2635 [hep-th].

\bibitem{Morales:1996bp}
  J.~F.~Morales and M.~Serone,
  Nucl.\ Phys.\ B {\bf 481} (1996) 389
  [hep-th/9607193].
  
\bibitem{Nakajima:2003uh}
  H.~Nakajima and K.~Yoshioka,
  ``Lectures on instanton counting,''
  arXiv:math/0311058.
  
\bibitem{Nakayama:2011be}
  Y.~Nakayama and H.~Ooguri,
  ``Comments on Worldsheet Description of the Omega Background,''
  Nucl.\ Phys.\  B {\bf 856}, 342 (2012)
  [arXiv:1106.5503 [hep-th]].

\bibitem{Nekrasov:2002qd}
  N.~A.~Nekrasov,
  ``Seiberg-Witten Prepotential From Instanton Counting,''
  Adv.\ Theor.\ Math.\ Phys.\  {\bf 7} (2004) 831
  [arXiv:hep-th/0206161].
 
\bibitem{Nekrasov:2003rj}
  N.~Nekrasov and A.~Okounkov,
  ``Seiberg-Witten theory and random partitions,''
  arXiv:hep-th/0306238. 
  
\bibitem{Ooguri:2002gx}
  H.~Ooguri and C.~Vafa,
  ``Worldsheet Derivation of a Large N Duality,''
  Nucl.\ Phys.\  B {\bf 641} (2002) 3
  [arXiv:hep-th/0205297].  
  
\bibitem{Seiberg:1994rs}
  N.~Seiberg and E.~Witten,
  ``Monopole Condensation, And Confinement In N=2 Supersymmetric Yang-Mills
  Theory,''
  Nucl.\ Phys.\  B {\bf 426} (1994) 19
  [Erratum-ibid.\  B {\bf 430} (1994) 485]
  [arXiv:hep-th/9407087].
  
\bibitem{Walcher:2007tp}
  J.~Walcher,
  ``Extended Holomorphic Anomaly and Loop Amplitudes in Open Topological
  String,''
  Nucl.\ Phys.\  B {\bf 817} (2009) 167
  [arXiv:0705.4098 [hep-th]].
  
\bibitem{Walcher:2007qp}
  J.~Walcher,
  Commun.\ Num.\ Theor.\ Phys.\  {\bf 3} (2009) 111
  [arXiv:0712.2775 [hep-th]].
  
\bibitem{Witten:1989ig}
  E.~Witten,
  ``ON THE STRUCTURE OF THE TOPOLOGICAL PHASE OF TWO-DIMENSIONAL GRAVITY,''
  Nucl.\ Phys.\  B {\bf 340} (1990) 281.

     
\end{thebibliography}
